\documentclass[aps,prb,amsmath,amssymb,superscriptaddress,nofootinbib,longbibliography,onecolumn]{revtex4-2}
\usepackage{graphicx}
\usepackage{mathrsfs}
\usepackage{bm}
\usepackage{textcomp,color}
\usepackage{xcolor}
\usepackage[colorlinks=true,urlcolor=blue,citecolor=blue,linkcolor=blue,bookmarks=false, pdfstartview={FitH}]{hyperref}
\usepackage[capitalize]{cleveref} 
\usepackage{amsmath}
\usepackage[normalem]{ulem} 

\newcommand{\beq}{\begin{equation}}
\newcommand{\eeq}{\end{equation}}
\newcommand{\bea}{\begin{eqnarray}}
\newcommand{\eea}{\end{eqnarray}}
\newcommand{\bse}{\begin{subequations}}
\newcommand{\ese}{\end{subequations}}
\newcommand{\nn}{\nonumber}
\newcommand{\bwt}{\begin{widetext}}
\newcommand{\ewt}{\end{widetext}}

\newcommand{\bk}{{\bf k}}
\newcommand{\bp}{{\bf p}}
\newcommand{\bq}{{\bf q}}
\newcommand{\br}{{\bf r}}

\newcommand{\bA}{{\bf A}}

\newcommand{\bl}{{\boldsymbol{\ell}}}

\begin{document}
	
\title{Electronic Raman scattering from 2D metals with broken inversion symmetry}
\author{Mojdeh Saleh}
\affiliation{Department of Physics, Concordia University, Montreal, QC H4B 1R6, Canada}
\author{Saurabh Maiti}
\affiliation{Department of Physics, Concordia University, Montreal, QC H4B 1R6, Canada}
\date{\today}

\begin{abstract}
Lack of inversion symmetry in metals breaks SU(2) symmetry which results in spin-splitting of the electronic states at the Fermi level due to various types of spin–orbit coupling (SOC) such as Dresselhaus, Rashba, or Ising (also called valley-Zeeman). This splitting is known to enable both incoherent spin-flip excitations and coherent chiral-spin modes. Another effect of breaking of SU(2) is the introduction of a \textit{direct} spin-photon interaction. We use this concept to formulate a theory of inelastic scattering of photons from the charge carriers of such a system [electronic Raman scattering (eRS)]. As a result of broken SU(2), we show that the eRS probe, unlike conventional theory of Raman scattering, couples to spin excitations even without tuning the laser to an internal resonance. We show that the spin-dependent excitations induced by photon scattering are sensitive to the polarization geometries as well as to the spin structure of the Hilbert space of the low-energy states. As a concrete realization, we examine doped/gated graphene on substrates with strong SOC with various compositions of Rashba and valley-Zeeman SOC and compare their spectra with those for a model 2D electron gas (2DEG). The spectra are shown to have a resonant feature in select polarization geometries near the SOC-splitting energy and, importantly, is shown to be different in the two systems. The signal in graphene systems is shown to be stronger than that in a 2DEG by orders of magnitude owing to the large Dirac velocity. We also outline how the lineshapes from the spectra can be used to infer various components of SOC in the system.

\end{abstract}

\maketitle

\tableofcontents

\section{Introduction}\label{Sec:Introduction}
Inelastic light scattering spectroscopy is a popular tool that measures the spectrum of the scattered photons from a sample allowing us to infer excitation properties of the system. In this context, Resonant Inelastic X-ray scattering (RIXS)~\cite{Schuelke2007} and Raman scattering~\cite{Weber2000} are two popular spectroscopic techniques to study the excitation properties of matter. In RIXS, the scattering  involves Hilbert space of core states that can be $\sim 100-1000$s of eV below the Fermi level. This probe is orbital sensitive and also transfers momenta that are comparable to the Brillouin zone (BZ) size. However, if one is interested in generic properties such as lattice/bond vibrations, or even the low-energy excitation properties of many-electron system, it is advantageous to work with Raman scattering. While it does not transfer significant momentum, it has higher energy resolution and only involves the Hilbert space within a few eVs of the Fermi level. 

Exploiting this, Raman scattering is extensively used to identify optical phonons and molecular vibrations. In correlated matter, this spectroscopy is used to probe low-energy spin excitations in magnetic materials, quasiparticle excitations in superconductors, density wave systems, and even partially polarized metals~\cite{Devereaux2007}. For phonon/molecular vibrations, one can use general group theory and classical ideas of polarizability of the medium to model the response. However, in the case of correlated \textit{electronic} systems it becomes necessary to understand the nature of the Hilbert space in which the light-matter interaction is to be modeled as it can induce additional selection rules~\cite{Benek-Lins2024}. 

In this work, we will be interested in the \textit{electronic} Raman scattering (eRS) from low-energy excitations of the many-electron metallic system in the absence of inversion symmetry. The lack of inversion lifts the SU(2) symmetry of the conduction and valence bands of the system through the introduction of a spin-orbit coupling (SOC) term of either the Dresselhaus type~\cite{Dresselhaus1955}, Rashba type~\cite{Bychkov1984} or Ising type~\cite{Kormanyos2014}. This splits the  doubly degenerate bands often imparting chirality to the spins of the eigenstates. Such a splitting at the Fermi level allows for incoherent spin-flip excitations as well as coherent chiral-spin excitations in the low-energy sector of the system. The latter, being coherent, are particularly interesting and have been detected in many materials: CdTe quantum wells~\cite{Perez2007,Baboux2013,Baboux2015,Perez2016,Karimi2017,maiti2017}, topological surface states of Bi$_2$Se$_3$~\cite{kung2017}, Moiré superlattice formed by WSe$_2$/WS$_2$ layers~\cite{Shan2025}. Interestingly, all of them used resonant eRS where the incoming laser was tuned to an internal resonance in the system.

There are a few subtle points here. The popular theory of eRS involves photon-matter interaction that does not couple to the spin degree of freedom. Nevertheless, if the material has atomic SOC (which preserves the SU(2) symmetry of bands), the spin excitations in such a system, no matter the origin, can be probed in certain polarization geometries of the experiment when the incoming photon is tuned to an internal resonance between bands (resonant eRS). No direct photon-spin interaction is needed~\cite{Shastry1990,Shastry1991}. It is in this context that previous studies were able to model chiral-spin excitations using resonant eRS~\cite{Chaplik2012,maiti2017}. 

Breaking of inversion symmetry reorganizes the low-energy Hilbert space of the system by lifting the SU(2) degeneracy of the bands. This has two distinct consequences. First, it modifies the electronic band structure  leading to the emergence of collective excitations called the chiral-spin modes, which have been extensively studied in recent years ~\cite{Maslov2022,shekhter2005,ashrafi2012,maiti2015a,maiti2015b,maiti2016,maiti2017,Kumar2021,Mojdeh2025}. Second, it alters the structure of the photon–matter interaction vertex and this received little attention until the surfacing of results from a recent study where a bare bulk plasmon was observed in resonant eRS from BiTeI crystal~\cite{Lee2024}. This was remarkable as plasmons are suppressed in the conventional theory of Raman scattering and are only indirectly detected when they hybridize with other Raman active modes. The observation was attributed to (i) modified photon-matter coupling due to the breaking of SU(2) in the full Hilbert space, an effect similar to singlet-triplet mixing in the absence of inversion, along with (ii) the presence of finite spin-charge susceptibility of the system [also due to breaking of SU(2)]~\cite{Sarkar2024a}. This was the first demonstration of a fundamentally different coupling of light to matter arising from breaking of inversion symmetry. Nevertheless, the modeling in this study was restricted to charge excitations and that too in the resonant limit of eRS. As a result, the fundamental (non-resonant) aspects of the photon-spin interaction due to SU(2) symmetry breaking is still not fully understood and this is precisely what this work targets. Since operating in resonance conditions effectively performs Hilbert space projections, from where an effective spin-photon interaction can emerge, to study a fundamental modification of photon-matter interaction it is necessary to study the non-resonant limit.

The non-resonant limit of eRS has been studied in the past in materials like monolayer and bilayer graphene both theoretically~\cite{Kashuba2009,Falko2010,Falko2012} and experimentally~\cite{Gallais2016,Gallais2019}, albeit, with inversion preserved. In general, non-resonant eRS becomes important for systems (such as graphene) where there aren't any eigenstates of the system that can resonate with the photons of typical Raman lasers. In this work, we show that the modification of the photon-matter coupling arising from breaking of inversion and the lifting of SU(2) introduces several interband terms in the Raman tensor and alters the scattering cross-section in ways that are characteristic of systems with different spin structure of the Hilbert space. The first system we study is doped/gated graphene on a transition metal dichalcogenide (TMD) / heavy metal substrate which induces Rashba and valley-Zeeman types of SOC in the graphene layer. This is a two-valley system with a $4\times4$ size of the Hilbert space. The second system is the same, but with energies far from the chemical potential projected out. This is a $2\times2$ Hilbert space. The photon-matter coupling in the two systems are shown to have different structures. This will be discussed in detail in Sec. \ref{Sec:erS_graphene}. While the projected system gets the right frequencies for spin excitations, we show that the differential cross-section is significantly weaker in the projected model, highlighting the relevance of higher energy bands in modeling the response. Further, SOC induces resonances and step jumps in all polarization geometries (XX, XY, RL) except RR, where X,Y refer to in-plane polarization directions and R,L refer to right/left circularly polarized light. The coherent resonance feature is shown to couple to the scattering cross-section only in the presence of Rashba-like SOC that couples momentum to spin.

We also show that the polarization dependence of the response is sensitive to the structure of the Hilbert space. To demonstrate this, we look into a model for the 2D electron gas (2DEG). While the low-energy spectrum (involving the spin split chiral states) is similar to the case of projected graphene model, the photon-matter coupling is different leading to a differential cross-section where RR geometry consists only of chiral excitations in the 2DEG case and only non-chiral ones in the graphene model.

The rest of the article is organized as follows. In Sec. \ref{Sec:CrossSection} we formulate the derivation of the cross-section in arbitrary Hilbert space. In Sec. \ref{Sec:erS_graphene} we discuss the eRS cross-section in SOC graphene and discuss the role of Rashba and VZ types of SOC. In Sec. \ref{Sec:Projection} we discuss the cross-section computed in a low-energy projected model. In Sec. \ref{Sec:Comparison_2DEG} we present results for a 2DEG with Rashba SOC to highlight the Hilbert space sensitivity of the cross-section. In Sec. \ref{Sec:Conclusion} we summarize the main findings and present an estimate for the size of the predicted effects in available systems. We point out that performing eRS on graphene like systems as opposed of 2DEGs presents a better likelihood of observing the predicted effects. Finally, in the appendix we present some details of the calculation of the differential cross-section.

\section{General scattering cross-section}\label{Sec:CrossSection}
We couple photons to our electronic system through the minimal coupling prescription. As a perturbative series in powers of the vector potential $\hat{\bA}$, we get\footnote{The conventional Raman scattering process involves a matrix element with photon states which only picks contributions from terms linear and quadratic in $\hat\bA$, and hence the Hamiltonian is expanded to $\mathcal O(A^2)$.}
\beq\label{eq:Hint_Def}
\hat{\mathcal H}[\bp,\hat \bA(\vec r,t)]=\hat{\mathcal{H}}_0(\bp)+\underbrace{e\hat v_\alpha(\bp)\hat A_{\alpha}(\vec r,t)+\frac{e^2}2\hat I_{\alpha\beta}(\bp)\hat A_{\alpha}(\vec r,t)\hat A_{\beta}(\vec r,t)}_{\equiv \hat{\mathcal H}_A(\vec r,t)}.
\eeq
where $\bp$ is the momentum quantum number, $-e$ is the electron charge, and the repeated indices (corresponding to in-plane components) are summed over. We have introduced $\hat{\mathcal H}_A$ to denote the perturbation of the original low-energy Hamiltonian $\hat{\mathcal H}_0$ due to coupling to the photon field. The bold vectors $\mathbf x$ denote in-plane vectors whereas the $\vec x$ vectors refer to vectors in 3D space. The linear-in-$\bf A$ term with the velocity operator $\hat v_\alpha(\bp)$ is the elementary photon-electron vertex and we call this the absorption vertex. The quadratic-in-$\bf A$ term is the diamagnetic coupling via the inverse mass tensor $\hat I_{\alpha\beta}(\bp)$. This prescription is valid for any size of the Hilbert space. If the low-energy Hamiltonian is analytic in $\bp$, then we have $\hat v_\alpha(\bp)=\partial_{p_\alpha}\hat{\mathcal{H}_0}(\bp)$ and $\hat I_{\alpha\beta}(\bp)=\partial_{p_{\alpha}}\partial_{p_{\beta}}\hat{\mathcal H}_0$. If the low-energy Hamiltonian is not analytic in $\bp$, then one needs to perform Hilbert space projections to arrive at the right matter-photon coupling. This will be discussed in Sec. \ref{Sec:Projection}.

The differential cross-section of scattering into a solid angle $d\mathcal O$ and within a frequency range $d\Omega$ is then given by~\cite{Devereaux2007}
\bea\label{eq:DiffXSec}
\frac{d^2\sigma}{d\mathcal{O}d\Omega}&=&\hbar r_0^2\frac{\Omega_{\rm S}}{\Omega_{\rm I}}\sum_{i,f}\frac{e^{-E_i/k_BT}}{\mathcal Z}|M_{fi}|^2\delta(E_f-E_i+\underbrace{\hbar\Omega_{\rm S}-\hbar\Omega_{\rm I}}_{\equiv-\hbar\Omega})\nn\\
&\stackrel{T\rightarrow0}{=}&\hbar r_0^2\frac{\Omega_{\rm S}}{\Omega_{\rm I}}\sum_{i,f}|M_{fi}|^2\delta(E_f-E_i-\hbar\Omega).
\eea
Here, $r_0\equiv e^2/4\pi\epsilon_0m_ec^2$ is the classical radius of the electron with mass $m_e$, $\Omega_{\rm S}$ and $\Omega_{\rm I}$ are the scattered (detected) and incident photon frequencies, respectively, $i,f$ label all possible initial and final states of the electronic system (consistent with the ground state and allowed excited state energies $E_i,E_f$), $\Omega=\Omega_{\rm I}-\Omega_{\rm S}$ is the Raman shift, and the dimensionless matrix element $M_{fi}$ is given by
\bea\label{eq:Mfi2}
\frac{M_{fi}}{m_e}=\left[-\sum_\nu\text{'}\left(\frac{\langle f|\hat v_\alpha \ell_{\alpha}^{\rm S*}|\nu\rangle\langle \nu|\hat v_{\alpha'} \ell_{\alpha'}^{\rm I}|i\rangle}{E_\nu-\hbar\Omega_{\rm I}-E_i}+\frac{\langle f|\hat v_\alpha \ell_{\alpha}^{\rm I}|\nu\rangle\langle \nu|\hat v_{\alpha'} \ell_{\alpha'}^{\rm S*}|i\rangle}{E_\nu+\hbar\Omega_{\rm S}-E_i}\right)+\ell_{\alpha}^{\rm S*}\ell_{\beta}^{\rm I}\langle f|\hat I_{\alpha\beta}|i\rangle\right]\delta_{\bk_f,\bk_i},
\eea
where $\sum$' denotes the restriction to sum only over vertical intermediate states, which is due to the small photon momentum (see Appendix \ref{Sec:App1}). Further, due to the $\delta_{\bk_f,\bk_i}$, the sum over the final states $f$ in the differential cross-section in Eq. \eqref{eq:DiffXSec} also gets reduced to only the vertical ones. The vectors $\boldsymbol{\ell}^{\rm I}, \boldsymbol{\ell}^{\rm S}$ denote the polarizations of incoming and scattered photons, respectively. The net contribution to the scattering matrix element consists of a direct (contact) contribution, which does not involve intermediate electronic states, and an indirect (two-step) contribution, which results in absorption/emission processes through intermediate electronic states $E_\nu$. The non-resonant limit is when the photon energy is larger than the eigen energies of the states in our Hilbert space, i.e. $\hbar\Omega_{\rm I}\gg E_\nu,E_i,\hbar\Omega$. In such a case, we can approximate Eq. \eqref{eq:Mfi2} as (see Appendix \ref{Sec:App1})
\bea\label{eq:NewMFi}
M_{fi}&=&\langle f|\hat M_{\alpha\beta}|i\rangle\ell_{\alpha}^{\rm S*}\ell_{\beta}^{\rm I}~\delta_{\bk_f,\bk_i},\nn\\
\text{such that }\frac{\hat M_{\alpha\beta}}{m_e}&\approx&\underbrace{\frac1{\hbar\Omega_{\rm I}}[\hat v_\alpha,\hat v_\beta]}_{\rm leading~indirect}\underbrace{-\frac{1}{\hbar^2\Omega^2_{\rm I}}\left(\hat v_\alpha[\hat v_\beta,\hat H_0]+\hat v_\beta[\hat v_\alpha,\hat H_0]\right)}_{\rm subleading~ indirect}+\underbrace{\hat I_{\alpha\beta}}_{\rm direct}.
\eea
Here, the indirect term is further expanded into the leading and subleading terms which are identified by the powers of $1/\hbar\Omega_{\rm I}$. In the conventional 1-band theory of Raman scattering, $\hat H_0$ is simply $\bp^2/2m$ with no matrix structure. As a result, the commutators of Eq. \eqref{eq:NewMFi} are zero leading to the absence of indirect processes and leaving us with just the direct one through the inverse mass term\footnote{This is what is known as the effective mass approximation.}. In larger Hilbert spaces, the matrices do not commute in general leading to a finite contribution from the indirect processes. The leading indirect contribution, sometimes called the $A_2$ response operator, takes the same form as derived in the context of orbital effects~\cite{Kashuba2009,Silva2020,Udina2026}. The final result for the differential cross-section (for the Stoke's process) can then be cast into the form:
\bea\label{eq:DiffXSec2}
\frac{d^2\sigma}{d\mathcal{O}d\Omega}
&\stackrel{T\rightarrow0}{=}&\hbar r_0^2\frac{\Omega_{\rm S}}{\Omega_{\rm I}}\sum_{i}\sum_{f[i]}|m_{\alpha\beta, f[i]i}\ell_{\alpha}^{\rm S*}\ell_{\beta}^{\rm I}|^2~n_{\rm F}(E_i)[1-n_{\rm F}(E_{f[i]})]~\delta(E_{f[i]}-E_i-\hbar\Omega),
\eea
where $m_{\alpha\beta,f[i]i}$ is the Raman tensor for the transition $i\rightarrow f[i]$: $\langle f[i]|\hat M_{\alpha\beta}|i\rangle$ and the sum over $f[i]$ is restricted to those final states which have $\bk_f=\bk_i$ (vertical bands). The creation and annihilation operators of $\hat M_{\alpha\beta}$ act on the fermionic occupation yielding the factor $n_{\rm F}(E_i)[1-n_{\rm F}(E_f)]$, where $n_{\rm F}(E)$ is the Fermi-Dirac distribution function~\cite{Devereaux2007,Blum1970}.

\subsection{Effect of inversion-breaking spin-orbit coupling}
While Eq. \eqref{eq:DiffXSec2} is general, the spin structure of the \textit{multiband} Hilbert space characteristically modifies the Raman tensor $\hat m_{\alpha\beta}$ (which is a matrix in the Hilbert space). The modification enters at three places: (i) implicitly in the SOC-renormalized electronic spectrum, leading to spin-split energies $E_\alpha$, (ii) the wavefunction $|\alpha\rangle$, and (iii) in the spin structure of the absorption vertex $\partial_{p_{\alpha}}\hat{\mathcal H}_0$. In the absence of inversion symmetry, the SU(2) symmetry of the Hilbert space is broken. This allows vectors (like momentum)  to  couple to pseudovectors (spin matrices). As a result of this coupling in the Hamiltonian, the velocity operator in \cref{eq:Hint_Def} and the wavefunctions acquire a spin-character which directly affects the scattering matrix elements. This leads to a \textit{direct} spin-photon interaction that would be absent in the presence of inversion symmetry. While the point (i) above leads to zero-field spin splitting and the existence of chiral-spin excitations~\cite{Maslov2022}, the points (ii) and (iii) affect how these excitations couple to the scattering cross-section. It is already known that such a direct spin-photon coupling leads to Electric dipole spin resonance (EDSR) in photon absorption (see ~\cite{Maslov2022} and references therein). Here, we explore its role in photon scattering in some popular model systems.

\section{Raman signal in gated spin-orbit coupled Graphene}\label{Sec:erS_graphene}
The case of monolayer graphene on a TMD/heavy metal substrate with only Rashba type SOC is a good \textit{multiband} model that describes the fundamentally different electronic Raman excitations that are induced due to inversion symmetry breaking. Such a model would be valid in, e.g., graphene/Ni(111)~\cite{Varykhalov2008}. The low-energy Hamiltonian and the coupling to the vector potential for such a system are written as~\cite{Kumar2021,Maslov2022}:
\bse
\bea
\hat H^{\rm G,R}&=&\Psi^\dagger(\hat{\mathcal H}^{\rm G,R}_0 + \hat{\mathcal H}^{\rm G,R}_A)\Psi,\\
\text{where}~\hat{\mathcal{H}}^{\rm G,R}_0&=& v_F(\tau^3\hat s^0\hat{\sigma}^1 p_x+\hat s^0\hat{\sigma}^2 p_y)+\frac{\lambda_{\rm R}}{2}(\tau^3\hat s^2\hat\sigma^1-\hat s^1\hat\sigma^2),\label{eq:modelG4Ra}\\
\hat{\mathcal{H}}^{\rm G,R}_A&=&\underbrace{v_F\tau^3\hat s^0\hat{\sigma}^1}_{\partial_{p_x}\hat{\mathcal H}_0} e\hat A_x(\vec r,t)+\underbrace{v_F\hat s^0\hat{\sigma}^2}_{\partial_{p_y}\hat{\mathcal H}_0} e\hat A_y(\vec r,t),\label{eq:modelG4Rb}
\eea
\ese
where the superscript ${\rm G}$ indicates Graphene, ${\rm R}$ indicates Rashba, $\tau^3=\pm1$ for the respective valleys of graphene, $\hat \sigma^i$ and $\hat s^i$, with $i\in\{1,2,3\}$, are the Pauli matrices in sublattice ($A,B$) and spin ($\uparrow,\downarrow$) spaces, respectively, and the basis of the Hamiltonian representation is $\Psi^\dagger=(c^\dagger_{\bp,\uparrow,A},c^\dagger_{\bp,\downarrow,A},c^\dagger_{\bp,\uparrow,B},c^\dagger_{\bp,\downarrow,B})$. Further, $\lambda_{\rm R}$, which has units of energy, denotes the Rashba coupling experienced by the graphene layer from the substrate. Since our objective is to study the effect of inversion breaking, we ignore trigonal warping in the above model. Such a term appears as $\sim p^2/2m_w$, with $m_w=3\gamma_0/v_F^2\sim 1.5 m_e$ ($\gamma_0\approx2.8~eV$ is the nearest neighbor coupling~\cite{Castro2009}). Due to this term, the indirect terms of $\mathcal{O}(m_ev^2_{\rm F}/\hbar\Omega_{\rm I})$ will be corrected by $\mathcal{O}(\mu m_e/\hbar\Omega_{\rm I}m_w)$ which is smaller by a factor of $\mu/m_wv_{\rm F}^2$ which is $\sim 1/75$ for $\mu\sim 100$meV. A direct contribution is also induced due to this trigonal warping term ~\cite{Kashuba2009} which is $\mathcal O(m_e/m_w)$, and this is $\hbar\Omega_{\rm I}/m_wv^2_{\rm F}\sim 1/5$ times the indirect contribution. Moreover, these terms do not induce a spin sensitive response and for these reasons, we will drop the trigonal warping terms going ahead. Finally, to ensure metallic behavior, we shall be interested in the case where the chemical potential lies well within the dispersive bands. Without loss of generality, we shall assume it to be the conduction bands. In an experiment, the control of the chemical potential can be achieved by gating or doping the setup.

\begin{figure}[htp]
    \centering
    \includegraphics[width=0.9\linewidth]{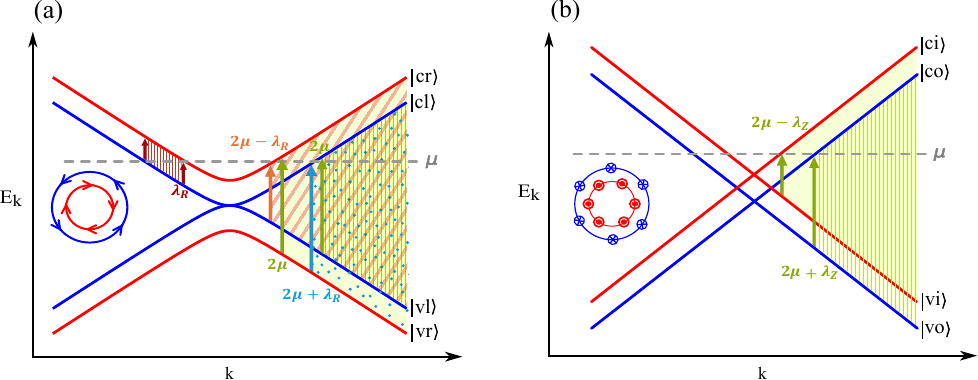}
    \caption{Electronic structure and possible excitations for systems in which (a) Rashba SOC and (b) valley--Zeeman SOC dominate. The insets show the chiral/spin texture (l--left, r--right, i--in, o--out) at the Fermi surface. In systems with dominant Rashba SOC, both chirality-preserving excitations at $2\mu$ and chirality-flipping excitations at $\lambda_{\rm R}$, $2\mu-\lambda_{\rm R}$, and $2\mu+\lambda_{\rm R}$ are allowed Raman excitations. These excitations correspond respectively to the green, red, orange, and blue arrows in panel (a), with shaded regions of matching colors indicating the associated continua. In contrast, when valley--Zeeman SOC dominates, only spin-preserving excitations appear in Raman at energies $\ge 2\mu \pm \lambda_{\rm Z}$. The arrows mark the threshold energies, while the shaded regions denote the continua of the same type of excitations.}
    \label{fig:BandStructure}
\end{figure}

The Hamiltonian in Eq. \eqref{eq:modelG4Ra} has the spectrum shown in Fig. \ref{fig:BandStructure}a where both the conduction and valence bands are split by $\lambda_{\rm R}$. Each band has a chirality that is also indicated in the figure. The figure also indicates the energies at which one may get chirality preserving vertical excitations (at energies $\ge2\mu$) and the chirality flipping ones (at energies $\ge2\mu\pm\lambda_{\rm R}$, and at energy $=\lambda_{\rm R}$). Below we proceed to calculate the expected Raman scattering spectrum from this electronic structure.

\subsection{Raman tensor in the Rashba dominated case}
To calculate the differential cross-section in Eq. \eqref{eq:DiffXSec2}, we need to sum the mod-square of the Raman tensor $m_{\alpha\beta,f[i]i}$ for all allowed pairs of states $i$ and $f[i]$. In Eq. \eqref{eq:DiffXSec2}, $f[i],\nu$ belong to the low-energy subspace involving the four bands (that comprise the two chiral conduction and the two chiral valence bands). Using the form of the eigenvectors of $\mathcal H_0^{\rm G,R}$ from Ref. ~\cite{Kumar2021} we first calculate the Raman tensor for $\alpha,\beta\in{x,y}$ which reflect the choices of the scattering geometries. Observe that, although absorption vertex $\partial_{p_\alpha}\hat{\mathcal H}_0$ has a trivial spin structure, the wavefunctions do not. As a result, the Raman tensor can (and do) get affected. Due to the lack of quadratic terms, we will only have indirect process contributions, which are separated into leading (L) and subleading (SL) contributions as
\beq\label{eq:split}
\hat m_{\alpha\beta}=\hat m^{\rm L}_{\alpha\beta} + \hat m_{\alpha\beta}^{\rm SL}.
\eeq
As we will see below, the leading contributions are either zero or SOC independent. Thus, to study the effect of the spin structure of the Hilbert space on the scattering, we need to keep the subleading term. The Raman tensor for various $\alpha,\beta\in\{x,y\}$ are presented in Eqs. (\ref{eq:G_Mxx}-\ref{eq:G_Mxy}), where for brevity we have introduced the dimensionless factors
\begin{align}\label{eq:deffacts}
f_{\rm R}&\equiv\frac{m_ev_F^2\lambda_{\rm R}}{(\hbar\Omega_{\rm I})^2},
 ~~~~~~~~~~~f_{\mu}\equiv\frac{m_ev_F^2\sqrt{4v_{\rm F}^2p^2+\lambda^2_{\rm R}}}{2(\hbar\Omega_{\rm I})^2},\nn\\
r_{\mu}& \equiv \frac{2v_{\rm F}p}{\sqrt{4v_{\rm F}^2p^2+\lambda^2_{\rm R}}},~\
r_{\rm R} \equiv \frac{2\lambda_{\rm R}}{\sqrt{4v_{\rm F}^2p^2+\lambda^2_{\rm R}}},\nn\\
c_\pm&\equiv2f_\mu\pm f_{\rm R}, ~~~~~~~~~~~~d_\pm \equiv2f_\mu\pm f_{\rm R}[1\mp r_{\rm R}].
\end{align}
The following matrices are written in the space: $(|vr\rangle,|vl\rangle,|cl\rangle,|cr\rangle)$
\begin{align}\label{eq:G_Mxx}
\hat m^{\rm L}_{xx}(\theta_p)&=0,\nn\\
    \hat m^{\rm SL}_{xx}(\theta_p)&=-\begin{pmatrix}
        -D_+&-2ir_\mu f_{\rm R}\sin2\theta_p&2\tau^3r_\mu f_{\rm R}\cos2\theta_p&-i\tau^3d_+\sin2\theta_p\\
        2ir_\mu f_{\rm R}\sin2\theta_p&-D_-&-i\tau^3d_-\sin2\theta_p&-2\tau^3r_\mu f_{\rm R}\cos2\theta_p\\
        2\tau^3r_\mu f_{\rm R}\cos2\theta_p&i\tau^3d_-\sin2\theta_p&D_-&-2ir_\mu f_{\rm R}\sin2\theta_p\\
        i\tau^3d_+\sin2\theta_p&-2\tau^3r_\mu f_{\rm R}\cos2\theta_p&2ir_\mu f_{\rm R}\sin2\theta_p&D_+
    \end{pmatrix},
\end{align}
such that $D_\pm=c_\pm-d_\pm\cos2\theta_p$. Here $\theta_p$ is the angle the $\bp$-vector makes with the $x$-axis. Further,
\begin{align}\label{eq:G_Myy}
\hat m^{\rm L}_{yy}(\theta_p)&=0,\nn\\
    \hat m^{\rm SL}_{yy}(\theta_p)&=\hat m^{\rm SL}_{xx}(\theta_p+\pi/2),
\end{align}
and 
\begin{align}\label{eq:G_Mxy}
\hat m^{\rm L}_{xy}(\theta_p)&=-\frac{m_ev_{\rm F}^2}{\hbar\Omega_{\rm I}}\begin{pmatrix}
    0&0&0&2i\tau^3\\
    0&0&2i\tau^3&0\\
    0&2i\tau^3&0&0\\
    2i\tau^3&0&0&0
\end{pmatrix},\nn\\
    \hat m^{\rm SL}_{xy}(\theta_p)&=-\begin{pmatrix}
        d_+\sin2\theta_p&2ir_\mu f_{\rm R}\cos2\theta_p&2\tau^3r_\mu f_{\rm R}\sin2\theta_p&i\tau^3d_+\cos2\theta_p\\
        -2ir_\mu f_{\rm R}\cos2\theta_p&d_-\sin2\theta_p&i\tau^3d_-\cos2\theta_p&-2\tau^3r_\mu f_{\rm R}\sin2\theta_p\\
        2\tau^3r_\mu f_{\rm R}\sin2\theta_p&-i\tau^3d_-\cos2\theta_p&-d_-\sin2\theta_p&2ir_\mu f_{\rm R}\cos2\theta_p\\
        -i\tau^3d_+\cos2\theta_p&-2\tau^3r_\mu f_{\rm R}\sin2\theta_p&-2ir_\mu f_{\rm R}\cos2\theta_p&-d_+\sin2\theta_p
    \end{pmatrix},\nn\\~\nn\\
    \hat m^{\rm L}_{yx}(\theta_p)&=-\hat m^{\rm L}_{xy}(\theta_p),\nn\\
    \hat m^{\rm SL}_{yx}(\theta_p)&=\hat m^{\rm SL}_{xy}(\theta_p).
\end{align}
The diagonal entries of the above matrices are of significance in the presence of significant disorder that induces a low frequency Drude-tail which can also be analyzed~\cite{Gallais2016b}. We will work in a relatively cleaner limit where the broadening is well with the elastic Rayleigh linewidth, and hence ignore these matrix elements.

Even before calculating the differential cross-section, the structure of these matrices help us deduce the electronic excitations in the system that couple to the response in various polarization geometries. A few general observations are: (i) The leading order response is independent of SOC and is only finite for the Raman tensor $\hat m_{xy}$. This is a known result for monolayer graphene ~\cite{Kashuba2009}. (ii) The subleading terms are anisotropic involving $d$-wave like form factors. This anisotropy will not be directly measurable as the scattering cross-section will involve an angular average of $\theta_p\in[0,2\pi)$. As a result, the XX and YY averages are the same, restoring the emergent $C_{\infty v}$ symmetry of the low-energy Hamiltonian in the observable spectrum. Nevertheless, the anisotropy of the Raman tensor opens a route to having possible interferences with electronic correlations that can overlap with the form-factors of the Raman tensor. We leave this topic for a future endeavor. Let us now investigate the nature of the electronic excitations captured by various choices of polarizations.

\paragraph*{Parallel linear polarization:}
This corresponds to having $\boldsymbol{\ell}^{\rm I}\parallel\boldsymbol{\ell}^{\rm S}$. Since our spectrum is isotropic, without loss of generality, we align our axis to the $\boldsymbol{\ell}^{\rm I}$. Thus, for this configuration, we have $\boldsymbol{\ell}^{\rm I}=\boldsymbol{\ell}^{\rm S}=(1,0)$, which only picks up the Raman tensor component $m_{xx}$. Contributions in this geometry only arise from subleading terms and contains excitations between all pairs of states. The strongest weight is for excitations between valence and conduction bands of the same chirality which exists already with $\lambda_{\rm R}\rightarrow 0$. These excitations have a threshold at $\hbar\Omega=2\mu$ (see \cref{fig:BandStructure}a) due to Pauli exclusion. The spin/chirality-flipping excitations are enabled in the presence of SOC. These are between the valence and conduction bands (which begin at $2\mu\pm\lambda_{\rm R}$) and also between the spin-split conduction bands (which are at $\hbar\Omega=\lambda_{\rm R}$).

\paragraph*{Crossed linear polarization:}
This corresponds to having $\boldsymbol{\ell}^{\rm I}\perp\boldsymbol{\ell}^{\rm S}$. Thus, for this configuration we have $\boldsymbol{\ell}^{\rm S}=(0,1)$, which only picks up the Raman tensor component $m_{yx}$. In this geometry, there are finite leading contributions arising from excitations between chirality preserving states starting from $\hbar\Omega=2\mu$. The structure of the subleading contributions are the similar to that in $\hat m_{xx}$ (with phase shifted angular factors) and hence weaker than the leading contribution by a power of $1/\hbar\Omega_{\rm I}$. 

\paragraph*{Same-circular polarization:}
Using circularly polarized light allows one to superpose the $xx,xy$ tensor elements. The set-up has $\boldsymbol{\ell}^{\rm I}=\boldsymbol{\ell}^{\rm S}=(1,i)/\sqrt 2$. This is referred to as RR geometry. This picks up the superposition $[m_{xx}+m_{yy} + i(m_{xy}-m_{yx})]/2$. As mentioned earlier, for finite Raman shifts we are only interested in the non-diagonal entries of the Raman tensor $\hat m_{\alpha\beta}$ matrices. Referring to such matrices as $\hat{\bar m}_{\alpha\beta}$ ($\bar m\equiv m-{\rm diag}(m)$), we observe from Eqs. (\ref{eq:G_Mxx}-\ref{eq:G_Mxy}) that $\hat{\bar m}_{xx}=-\hat{\bar m}_{yy}$, and $\hat{\bar m}^{\rm SL}_{xy}=\hat{\bar m}^{\rm SL}_{yx}$. This results in the RR contribution to be $\propto(\hat m^{\rm L}_{xy}-\hat m^{\rm L}_{yx})/2=\hat m^{\rm L}_{xy}$ which only causes chirality preserving excitations and thus remains insensitive to the inversion breaking of the system. Thus, no spin splitting should be observed in this geometry.

\paragraph*{Counter-circular polarization:}
In this scattering geometry we have $\boldsymbol{\ell}^{\rm I}=\boldsymbol{\ell}^{\rm S*}=(1,i)/\sqrt 2$ and one refers to this as RL. This picks up $[m_{xx}-m_{yy}+i(m_{xy}+m_{yx})]/2$ combination. Using the properties of $\hat m_{\alpha\beta}$ tensors, at finite Raman shifts, this results in $\hat{\bar m}^{\rm SL}_{xx}+i\hat{\bar m}^{\rm SL}_{xy}$. The mod-square of these Raman tensors essentially sums up the (sub-leading) responses from the XX and XY geometries. Thus, the XX, XY and RL geometries will essentially contain the same information, with the XY carrying the additional leading contribution.

\subsection{Differential cross-section}\label{subsec:diffXSec}
To compute the differential cross-section, one simply needs to integrate the modulus-squared of the total Raman tensor for a polarization geometry over all initial states. The domain of initial states is the entire BZ, but the Fermi functions in Eq. \ref{eq:DiffXSec2} ensure the summation to be restricted to states respecting Pauli's exclusion rule. This summation over $i$ thus includes the sum over valleys of the system. The summation over Pauli-restricted states is further restricted to an angular average at $p\sim \hbar\Omega/v_{\rm F}$ due to the energy-conserving $\delta$-function in Eq. \eqref{eq:DiffXSec2}. In all the following calculations, we place the chemical potential, $\mu$, well within the conduction band so that $\mu\ge\lambda_{\rm R}$. The result of the computation for an arbitrary polarization geometry is 
\bea\label{eq:GR_diff}
\dfrac{d^2\sigma}{d\mathcal Od{\Omega}}&=&\frac{r_0^2}{\Omega_{\rm I}}\frac{S(m_ev_F)^2}{\pi\hbar^2}\times\nn\\
&&\left[\frac{2\Omega}{\Omega_{\rm I}}\Theta(\hbar\Omega-2\mu)~\Xi_{\rm L}~~+\right.\nn\\
&&~~~\left.\left\{s_1\big[\Theta(\hbar\Omega-2\mu-\lambda_{\rm R})+\Theta(\hbar\Omega-2\mu+\lambda_{\rm R})\big]+s_2\Theta(\hbar\Omega-2\mu)+s_3\delta(\lambda_{\rm R}-\hbar\Omega)\right\}~\Xi_{\rm SL}\right]\nn\\~\nn\\
\text{where }s_1&\equiv&\frac{\lambda_{\rm R}^2(\hbar^2\Omega^2-\lambda_{\rm R}^2)}{2\hbar^4\Omega\Omega_{\rm I}^3},\nn\\
s_2&\equiv&\frac{\lambda_{\rm R}(\lambda_{\rm R}^2+2\lambda_{\rm R}\hbar\Omega-\hbar^2\Omega^2)\text{sgn}(\lambda_{\rm R}-\hbar\Omega)}{4\hbar^3\Omega^3_{\rm I}}-\frac{2\lambda_{\rm R}^4+2\lambda_{\rm R}^3\hbar\Omega+3\lambda_{\rm R}^2\hbar^2\Omega^2-4\lambda_{\rm R}\hbar^3\Omega^3+\hbar^4\Omega^4}{8(\lambda_{\rm R}-\hbar\Omega)\hbar^3\Omega_{\rm I}^3}\nn\\
&&~~~+\frac{(-2\lambda_{\rm R}^2+\lambda_{\rm R}\hbar\Omega+\hbar^2\Omega^2)^2}{8(\lambda_{\rm R}+\hbar\Omega)\hbar^3\Omega^3_{\rm I}},\nn\\
s_3&\equiv&\frac{2\lambda_{\rm R}^3}{\hbar^3\Omega_{\rm I}^3}\bigg(\mu- \frac{\lambda_{\rm R}}{4}\log \frac{2\mu+\lambda_{\rm R}}{2\mu-\lambda_{\rm R}}\bigg),
\eea
where $$\Xi_{\rm L}\equiv|\bl^{\rm I}\times{\bl}^{\rm S*}|^2,~\Xi_{\rm SL}\equiv1+(\bl^{\rm I}\times\bl^{\rm I*})\cdot({\bl^{\rm S}}\times{\bl}^{\rm S*}),$$ and $S$ is the area of the sample exposed to the incoming beam. The polarization factors $\Xi_{\rm L},\Xi_{\rm SL}$ were also found in ~\cite{Kashuba2009}, where the subscripts L, SL label the leading or subleading nature of their origin. The overall scale of the cross-section is $\propto$ three factors: (i) A universal $r_0^2/\Omega_{\rm I}$ that is material independent; (ii) $S(m_ev_F/\hbar)^2$ that is only dependent on the Dirac-velocity and the sample area; and (iii) polarization factors. This last factor is model dependent but can broadly be further classified into terms $\propto\Omega/\Omega_{\rm I}$ or $(\Omega/\Omega_{\rm I})^3$ depending on the leading or subleading nature of the contribution, where $\Omega$ is the Raman shift with interesting features around the values $\in\{\lambda_{\rm R}, 2\mu, 2\mu\pm\lambda_{\rm R}\}$. 

\begin{figure}[htp]
    \centering
    \includegraphics[width=0.9\linewidth]{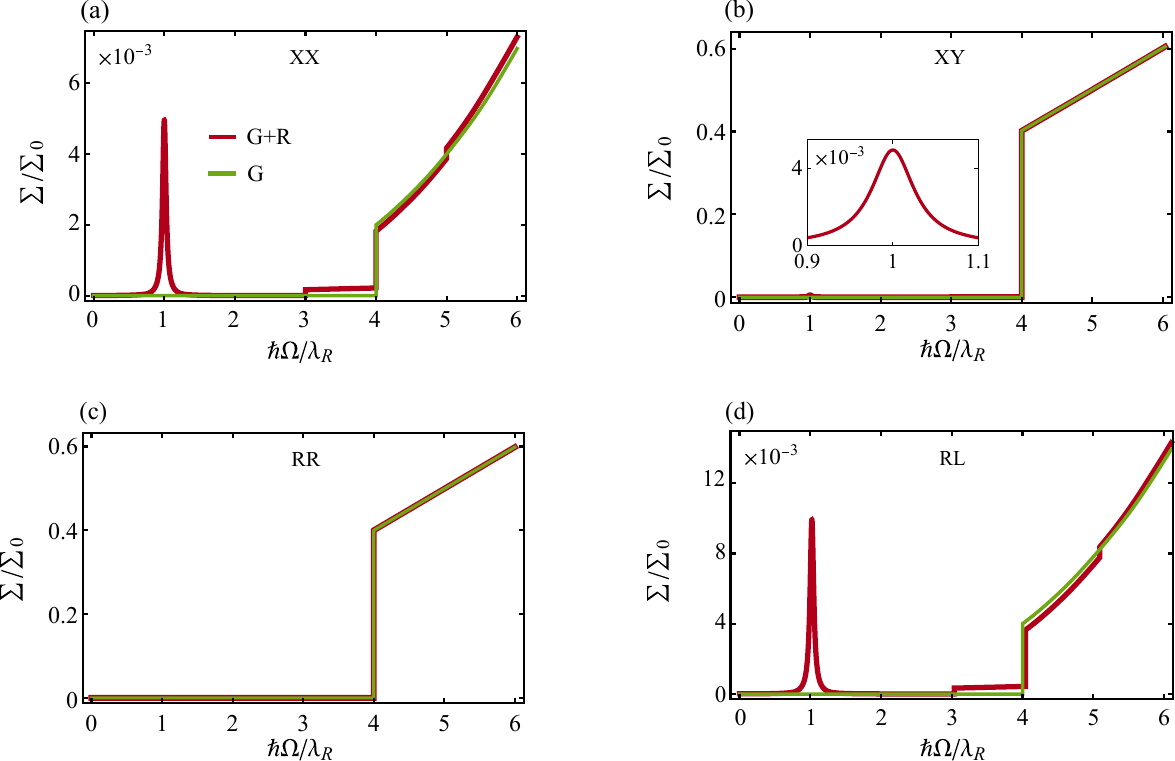}
    \caption{Differential scattering cross-section, $\Sigma\equiv d^2\sigma/d\mathcal Od\Omega $, normalized to $\Sigma_0\equiv r_0^2 S(m_ev_F/\hbar)^2/\pi\Omega_{\rm I}$ for graphene with proximity induced Rashba SOC(red) for (a) XX, (b) XY, (c) RR, and (d) RL scattering geometries. For comparison, we also show the cross-section for graphene without SOC (light green). The XY geometry has spin split features but is dominated by the spin-blind leading contribution. The inset shows a zoom-in around the region $\hbar\Omega\approx \lambda_{\rm R}$. The height of the peak is similar to that in the XX polarization. The RR geometry does not couple to spin-flip excitations. Here, $\mu=2\lambda_{\rm R},\hbar\Omega_{\rm I}=20\lambda_{\rm R}$. We included damping ($1/\tau$) to broaden the $\delta$-peak by setting $\lambda_{\rm R}\tau/\hbar=33$.}
    \label{fig:diff_cross_G_Rashba}
\end{figure}

The result for the differential cross-section for the four cases of polarization geometries discussed above are plotted in Fig. \ref{fig:diff_cross_G_Rashba}. All the plots have the same scale and thus the relative strengths of scattering can be compared directly. The first thing to note is that the XY and RR cross-sections are larger. This is due to the presence of the dominant leading contribution discussed above. Second, in line with the discussion above, the features due to spin splitting (peak at $\lambda_{\rm R}$ and steps at $2\mu\pm\lambda_{\rm R}$) are present in all geometries except in RR where they are identically zero. It can be verified explicitly from Eq. \ref{eq:GR_diff} that the responses in the various polarization geometries are such that XX+XY=RR+RL.

The sharp resonance we see at $\hbar\Omega=\lambda_{\rm R}$ coincides with interband excitation $|cl\rangle\leftrightarrow|cr\rangle$. We comment here that if one were to account for electronic correlations, this resonance would actually happen at the renormalized frequency of the chiral-spin modes\footnote{A precise modeling of the interaction effects will be presented in a future work.}. To understand how this can be brought about, note the presence of the constant, $\sin2\theta_p$ and $\cos2\theta_p$ form factors for the Raman tensor components. When we include electronic correlations, their matrix elements will involve a momentum transfer $\bp-\bp'$ corresponding to the electron-electron scattering process. This can be expanded in harmonics $\{\cos n\theta_p,\sin n\theta_p\}$. In general, each harmonic is capable of inducing a collective mode~\cite{kumar2017}. In this case, it is the interactions with the form factors of a constant and $\cos2\theta_p,~\sin2\theta_p$ that will couple to the leading and subleading parts, respectively, of the Raman tensors. However, for this article we shall continue in the non-interacting limit as we are more interested in developing the fundamental understanding of the role of inversion breaking on light-matter coupling.

\subsection{Differential cross-section in Valley-Zeeman (VZ) dominated case}
\begin{figure}[htp]
    \centering
    \includegraphics[width=0.9\linewidth]{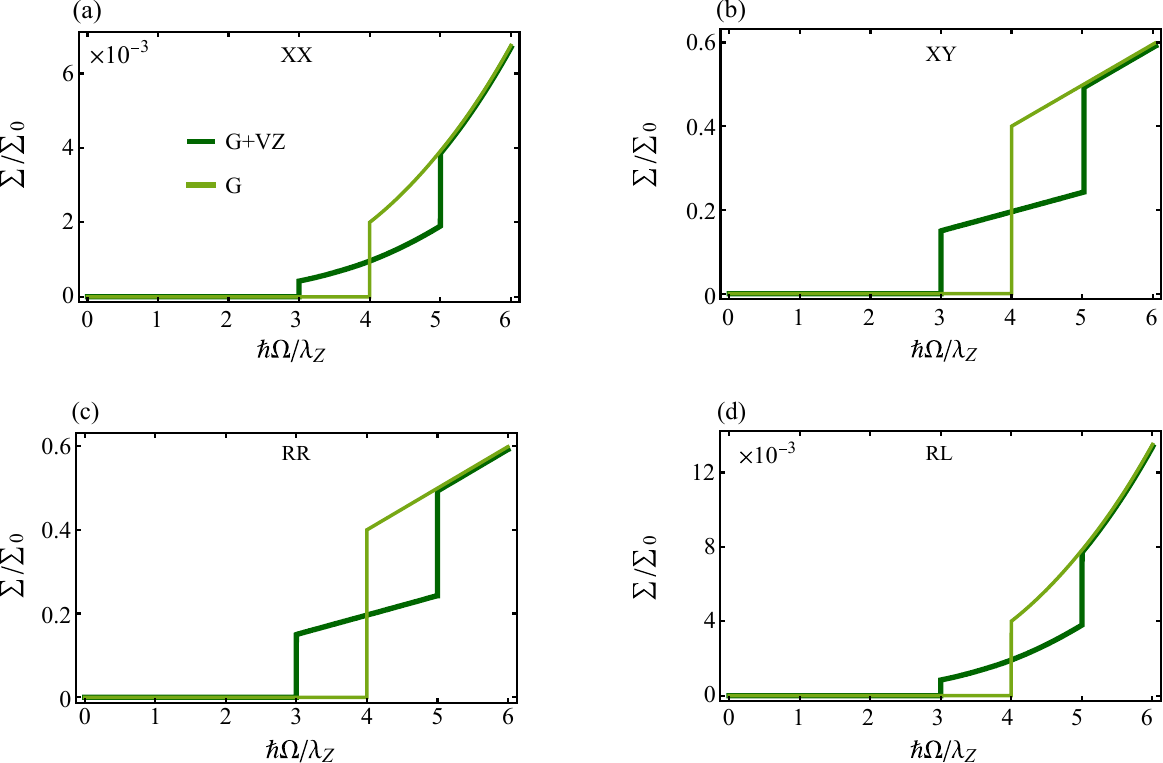}
    \caption{Normalized differential scattering cross-section ($\Sigma/\Sigma_0$, see caption to Fig. \ref{fig:diff_cross_G_Rashba}) for graphene with proximity induced VZ SOC (dark green line) for (a) XX, (b) XY, (c) RR, and (d) RL. For comparison, we also show the cross-section for graphene without SOC in light green. All features arise from spin-non-flip excitations. Although the splitting is seen here at $2\mu\pm\lambda_{\rm Z}$, it is not due to spin-flip excitations but due to the shift in the occupied/unoccupied energies of $i,o$ bands.  Here, $\mu=2\lambda_{\rm Z}, \hbar\Omega_{\rm I}=20\lambda_{\rm Z}$.}
    \label{fig:diff_cross_G_VZ}
\end{figure}

The results above were for the case for Rashba type of SOC. For systems where valley-Zeeman SOC dominates (such as in graphene/$\rm WS_2$~\cite{Wang2015}), the Hamiltonian and the vector potential perturbation, expressed in the same basis as Eqs. (\ref{eq:modelG4Ra},\ref{eq:modelG4Rb}) are given by 
\bse
\bea
\hat{\mathcal{H}}^{\rm G,VZ}_0&=& v_F(\tau^3\hat s^0\hat{\sigma}^1 p_x+\hat s^0\hat{\sigma}^2 p_y)+\frac{\lambda_{\rm Z}}{2}\tau^3\hat s^3\hat\sigma^0,\label{eq:modelG4Za}\\
\hat{\mathcal{H}}^{\rm G,VZ}_A&=&\hat{\mathcal{H}}^{\rm G,R}_A,\label{eq:modelG4Zb}
\eea
\ese
where $\lambda_{\rm Z}$ is the valley-Zeeman SOC in units of energy. The electronic structure and the spin texture at the Fermi surface are shown in Fig. \ref{fig:BandStructure}b. The $\bA$-perturbation is the same for both Rashba and valley-Zeeman cases because these are independent of $\bp$ and thus the $\bA$ field correction which only affects the $\bp$-dependent terms remains the same. This means that the $\hat M_{\alpha\beta}$ operators also remain the same as in the Rashba case and are blind to spin. However, unlike the Rashba case, the wavefunctions for $\mathcal H_0^{\rm G,VZ}$ (see Ref. ~\cite{Kumar2021}) are separable into spin up and down. As a result, the two spin states independently contribute to the Raman tensors $\hat m_{\alpha\beta}$ without any spin-flips. Indeed, applying Eq. \eqref{eq:NewMFi} to compute $\hat m_{\alpha\beta}$ in the space $\{|vo\rangle,|vi\rangle,|co\rangle,|ci\rangle\}$ we get
\begin{align}\label{eq:VZ_Mab}
\hat m^{\rm L}_{xx}(\theta_p)&=0,\nn\\
    \hat m^{\rm SL}_{xx}(\theta_p)&=-\frac{m_ev_F^3p}{(\hbar\Omega_{\rm I})^2}\begin{pmatrix}
        -4\sin^2\theta_p&0&-2i\tau^3\sin2\theta_p&0\\
        0&-4\sin^2\theta_p&0&-2i\tau^3\sin2\theta_p\\
        2i\tau^3\sin2\theta_p&0&4\sin^2\theta_p&0\\
        0&2i\tau^3\sin2\theta_p&0&4\sin^2\theta_p
    \end{pmatrix},\nn\\~\nn\\
    \hat m^{\rm L}_{yy}(\theta_p)&=0,\nn\\
    \hat m^{\rm SL}_{yy}(\theta_p)&=\hat m^{\rm SL}_{xx}(\theta_p+\pi/2),\nn\\~\nn\\
\hat m^{\rm L}_{xy}(\theta_p)&=-\frac{m_ev_{\rm F}^2}{\hbar\Omega_{\rm I}}\begin{pmatrix}
    0&0&2i\tau^3&0\\
    0&0&0&2i\tau^3\\
    2i\tau^3&0&0&0\\
    0&2i\tau^3&0&0
\end{pmatrix},\nn\\
    \hat m^{\rm SL}_{xy}(\theta_p)&=-\frac{m_ev_{\rm F}^3p}{(\hbar\Omega_{\rm I})^2}\begin{pmatrix}
        2\sin2\theta_p&0&2i\tau^3\cos2\theta_p&0\\
        0&2\sin2\theta_p&0&2i\tau^3\cos2\theta_p\\
        -2i\tau^3\cos2\theta_p&0&-2\sin2\theta_p&\\
        0&-2i\tau^3\cos2\theta_p&0&-2\sin2\theta_p
    \end{pmatrix},\nn\\~\nn\\
    \hat m^{\rm L}_{yx}(\theta_p)&=-\hat m^{\rm L}_{xy}(\theta_p),\nn\\
    \hat m^{\rm SL}_{yx}(\theta_p)&=\hat m^{\rm SL}_{xy}(\theta_p).
\end{align}
The relations between $\hat m_{xx}, \hat m_{yy}, \hat m_{xy},\hat m_{yx}$ are similar to the Rashba case (which is also the case with no SOC). The structure of the $\hat m_{\alpha\beta}$ makes it clear that only those excitations are possible that preserve the spin flavor ($o,i$) which in contrast to the Rashba scenario. Note that these Raman tensors are explicitly independent of $\lambda_{\rm Z}$. This is consistent with our expectation above that the response should consist of independent contributions from the two spin states. In fact, the only aspect making the response different from the case with no SOC is the energy of the excitations. Without SOC, the threshold for the interband excitation is at $2\mu$. In the presence of VZ SOC, the spin-preserving interband excitations begin instead at $2\mu\pm\lambda_{\rm Z}$ (see Fig. \ref{fig:BandStructure}b). Comparing \cref{eq:VZ_Mab} with Eqs. (\ref{eq:G_Mxx}-\ref{eq:G_Mxy}), it is also evident that the spin-flavor preserving excitations $|vx\rangle\leftrightarrow|cx\rangle$ for $x\in$ either $\{r,l\}$ or $\{o,i\}$ have the same structure and anisotropy. As a result, the differential cross-section will have threshold frequencies at $2\mu\pm\lambda_{\rm Z}$, but none at $2\mu$ itself:
\bea\label{GVZ_diff}
\dfrac{d^2\sigma}{d\mathcal Od\Omega}&=&\frac{r_0^2}{\Omega_{\rm I}}\frac{S(m_ev_F)^2}{\pi\hbar^2}\left(\frac{\Omega}{\Omega_{\rm I}}\Xi_{\rm L}+\frac{\Omega^3}{8\Omega^3_{\rm I}}\Xi_{\rm SL}\right)\big[\Theta(\hbar\Omega-2\mu-\lambda_{\rm Z})+\Theta(\hbar\Omega-2\mu+\lambda_{\rm Z})\big].
\eea
The differential cross-section is shown in \cref{fig:diff_cross_G_VZ}. Comparing to the Rashba case, we see the absences of the sharp resonance in the excitation spectrum and the step at $2\mu$, both of which would have flipped the spin (Fig. \ref{fig:BandStructure}b). The lack of the step at $2\mu$ is replaced by the spin-non-flip excitations at $2\mu\pm\lambda_{\rm Z}$.

\begin{figure}[htp]
    \centering
    \includegraphics[width=0.9\linewidth]{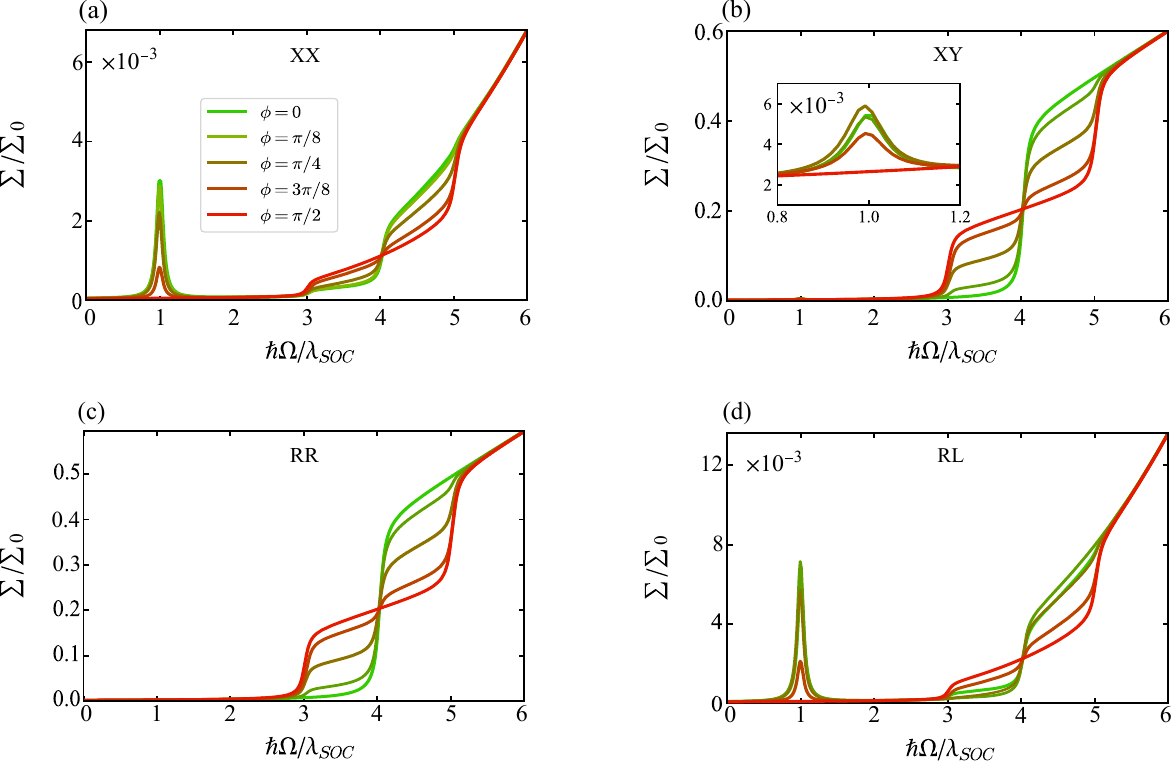}
    \caption{Normalized differential scattering cross-section for graphene with both Rashba and VZ SOCs for (a) XX, (b) XY, (c)RR, and (d) RL scattering geometries. The splitting in the spectrum is $\lambda_{\rm SOC}=\sqrt{\lambda_{\rm R}^2+\lambda_{\rm Z}^2}$. The parameter $\phi\equiv\arctan(\lambda_{\rm Z}/\lambda_{\rm R})$ tracks the contribution to the splitting from the two SOCs. It is evident that the weight of the resonance at $\lambda_{\rm SOC}$ is controlled by $\lambda_{\rm R}$, vanishing when $\lambda_{\rm R}\rightarrow 0$. The step jump at $2\mu$ is also controlled by dominance of the $\lambda_{\rm R}$, while the splitting around $2\mu$ is controlled by $\lambda_{\rm Z}$. Here,  $\mu=2\lambda_{\rm SOC}, \hbar\Omega_{\rm I}=20\lambda_{\rm SOC}$. We included a broadening by setting $\lambda_{\rm SOC}\tau/\hbar=20$. This leads to a tail in the leading term that extends to $\Omega\sim 0$ causing a background $\sim2\times10^{-3}$.}
    \label{fig:diff_cross_G_RashbaVZ}
\end{figure}

\subsection{Systems with both VZ and Rashba SOC} The case of VZ SOC shows that it is essential to mix the spin states in order to illicit a spin-flip excitation due to light. This is not possible unless the SOC couples the in-plane spins to momentum. In the more general case, monolayer graphene on a substrate will have both VZ and Rashba SOC. The Hamiltonian is given by including both $\lambda_{\rm R}$ and $\lambda_{\rm Z}$ dependent terms from Eqs. \eqref{eq:modelG4Ra} and \eqref{eq:modelG4Za} in the Hamiltonian. The eigenvalues of the system indicate that the spin splittings involve the parameter $\lambda_{\rm SOC}=\sqrt{\lambda_{\rm R}^2+\lambda_{\rm Z}^2}$. The various interband excitations $\in\{\lambda_{\rm SOC},2\mu,2\mu\pm\lambda_{\rm SOC}\}$. The Rashba contribution to the SOC splitting can be tracked by introducing another parameter $\phi$ such that $\tan\phi\equiv\lambda_{\rm Z}/\lambda_{\rm R}$. Changing the parameter $\phi$ would mean that we keep the SOC-splitting constant and play with its composition in terms of the two components.

The analytical calculation for this case is not tractable, but the numerical results for the differential cross-section computed form Eq. \eqref{eq:DiffXSec2} are shown in Fig. \ref{fig:diff_cross_G_RashbaVZ}. The various features of these plots are easily understandable in the context of the individual results for $\lambda_{\rm R}$ and $\lambda_{\rm Z}$:
\begin{itemize}
    \item The resonance due to spin-flip excitations between the conduction bands happens at the spin-splitting energy, $\lambda_{\rm SOC}$, but the spectral weight is controlled entirely by the presence of $\lambda_{\rm R}$ such that the weight vanishes when $\lambda_{\rm R}\rightarrow0~(\phi\rightarrow\pi/2)$.
    \item At higher energies there is a 3-step feature. The steps in the signal at $2\mu\pm\lambda_{\rm SOC}$ is dominated by the VZ SOC, while the step at $2\mu$ is a signature of presence of Rashba SOC.
    \item The strongest Raman signals appear in the XY and RR geometries, where the three-step feature can be used to extract the relative strengths of the two SOCs.
\end{itemize}

\section{Effective low-energy model and role of intermediary bands}\label{Sec:Projection}
In systems with valence and conduction bands that are also spin-split, there are high-energy excitations ($\hbar\Omega\gtrsim2\mu$) between the valence and conduction bands, and lower energy excitations ($\hbar\Omega\sim\lambda_{\rm SOC}$) between the spin split states of either the conduction or the valence bands. In cases when $\mu\gg\lambda_{\rm SOC}$ one is usually interested in the sharp low-energy excitations compared to the broad high-energy one. One can then move to the Hilbert space of just the spin-split bands by projecting out the other states using L{\"o}wdin's method~\cite{Lowdin1951}. The projection is done \textit{after} coupling the gauge field $\bA$ in the larger Hilbert space where the Hamiltonian is analytic. Performing this step with Rashba SOC yields~\cite{Kumar2021}
\bse
\bea
\hat{\mathcal{H}}^{\rm proj-G,R}_0&=&v_Fp\hat s^0+\frac{\lambda_{\rm R}}{2}(\bp_u\times\hat{\mathbf s})\cdot z,\label{eq:modelG2Ra}\\
\hat{\mathcal{H}}^{\rm proj-G,R}_A&=&v_F \bp_u\cdot e\hat{\bA}(\mathbf{r},t)\hat{s}^0+\frac{\lambda_{\rm R}}{2p}(e\hat{\bA}\times\bp_u)\cdot \hat z(\bp_u\cdot\hat {\mathbf s}),\label{eq:modelG2Rb}
\eea
\ese
where $\bp_u$ is the unit vector along $\bp$. This system has the same character of the eigenvectors and the energies as the $cl$-$cr$ sector of the $4\times4$ system. To apply Eq. \eqref{eq:NewMFi} to projected systems, the velocity operator in Eq. \eqref{eq:NewMFi} is identified with $\partial_{A_\alpha}\hat{\mathcal H}_A/e$. In fact, this is the correct identification. It is just that in Hamiltonians that are analytic, $\partial_{A_\alpha}\hat{\mathcal H}_A=\partial_{p_\alpha}\hat{\mathcal H}_0$. In projected systems, we have to find $\hat{\mathcal H}_A$ by projecting out the unwanted Hilbert space \textit{in the presence of} $\hat{\bA}$ and then expanding in $\hat \bA$. This is not the same as performing $\hat{\mathcal H}^{\rm proj}_0(\bp)\rightarrow\hat{\mathcal H}^{\rm proj}_0(\bp+e\bA)$. In general, the size of the Hilbert space in which $\mathbf A$ should be introduced is the one that captures all the states that can be reached with the photon, even if one is interested only in the low-energy excitations of the electronic system. This will ensure that the model for the light-matter interaction will capture all the processes that are necessary to model the response. An example of this is shown below.

We can calculate the following Raman tensors in the projected subspace $\{|cl\rangle,|cr\rangle\}$ using Eqs. \eqref{eq:modelG2Rb} and \eqref{eq:NewMFi}:
\begin{align}\label{eq:Proj M}
\hat m^{\rm L}_{xx}(\theta_p)&=0,\nn\\
\hat m^{\rm SL}_{xx}(\theta_p)&=-\frac{m_ev_{\rm F}\lambda_{\rm R}^2}{2p(\hbar\Omega_{\rm I})^2}\begin{pmatrix}
    -\frac{\lambda_{\rm R}\sin^2\theta_p}{v_{\rm F}p}&i\sin2\theta_p\\
    i\sin2\theta_p&\frac{\lambda_{\rm R}\sin^2\theta_p}{v_{\rm F}p}
\end{pmatrix},\nn\\~\nn\\
\hat m^{\rm L}_{yy}(\theta_p)&=0,\nn\\
\hat m^{\rm SL}_{yy}(\theta_p)&=\hat m^{\rm SL}_{xx}(\theta_p+\pi/2),\nn\\~\nn\\
\hat m^{\rm L}_{xy}(\theta_p)&=0,\nn\\
\hat m^{\rm SL}_{xy}(\theta_p)&=-\frac{m_ev_{\rm F}\lambda_{\rm R}^2}{2p(\hbar\Omega_{\rm I})^2}\begin{pmatrix}
    \frac{\lambda_{\rm R}\sin2\theta_p}{2v_{\rm F}p}&-i\cos2\theta_p\\
    -i\cos2\theta_p&-\frac{\lambda_{\rm R}\sin2\theta_p}{2v_{\rm F}p}
\end{pmatrix},\nn\\~\nn\\
\hat m^{\rm L}_{yx}(\theta_p)&=0,\nn\\
\hat m^{\rm SL}_{yx}(\theta_p)&=\hat m^{\rm SL}_{xy}(\theta_p).
\end{align}
Comparing these $\hat m$-tensors to the $cl-cr$ section of the full Hilbert-space system, we observe two important features of working in the projected system. First, the anisotropy of the Raman tensors are correctly captured\footnote{The diagonals terms are not of interest to us as they correspond to zero Raman shifts.}. Second, the overall scale of the Raman tensor in the projected case is reduced by $\lambda_{\rm R}/vp$. The source of this drop can be understood if we parse the individual contributions to the Raman tensor modeled in the full Hilbert space into those arising from individual intermediate states. In the projected system, there are no other intermediate states than those in the projected Hilbert space. In the full system, there are contributions from the intermediate states corresponding to the \textit{discarded} states. This is an important observation as it is customary to drop the higher Hilbert space states on grounds of them not contributing significantly to lower energies. This is an example where, even if the system is not resonating with the higher energy states, the latter contribute significantly to the spectral weight of the scattering process. Such non-resonant contributions are often referred to as `off-the-mass-shell' contributions. This is a field theory term that simply refers to a virtual state of a band that is at an energy different from the actual energy of the band. Such contributions always exist in quantum mechanics and their weights are further enhanced by finite lifetime effects. 

The differential cross-section calculated for the projected system from Eq. \eqref{eq:DiffXSec2} is:
\beq\label{eq:PGR_diff2}
\dfrac{d^2\sigma}{d\mathcal Od\Omega }=\frac{r_0^2}{\Omega_{\rm I}}\frac{S(m_ev_F)^2}{\pi\hbar^2}\left[\frac{\lambda_{\rm R}^3}{4\hbar^3\Omega^3_{\rm I}}~\text{arctanh}\frac{\lambda_{\rm R}}{2\mu}~\lambda_R\delta(\hbar\Omega-\lambda_{\rm R})~\Xi_{\rm SL}\right].
\eeq
The result should be compared with the $s_3$ term in Eq. \eqref{eq:GR_diff} which is the corresponding response in the full Hilbert space. $s_3$ has two contributions. One is $\mathcal{O}(\mu\lambda^3_{\rm R}/\Omega_{\rm I}^3)$, while the other is $\mathcal{O}(\lambda^5_{\rm R}/\mu\Omega_{\rm I}^3)$ which is $\lambda_{\rm R}^2/\mu^2$ (or $\lambda_{\rm R}^2/v_F^2p^2$) smaller. This second contribution to $s_3$ arises precisely from the $\lambda_{\rm R}/vp$ factor discussed above that arises from the $\{|cl\rangle,|cr\rangle\}$ subspace, and shows up explicitly in Eq. \eqref{eq:PGR_diff2} for the projected system. 

\begin{figure}[htp]
    \centering
    \includegraphics[width=0.9\linewidth]{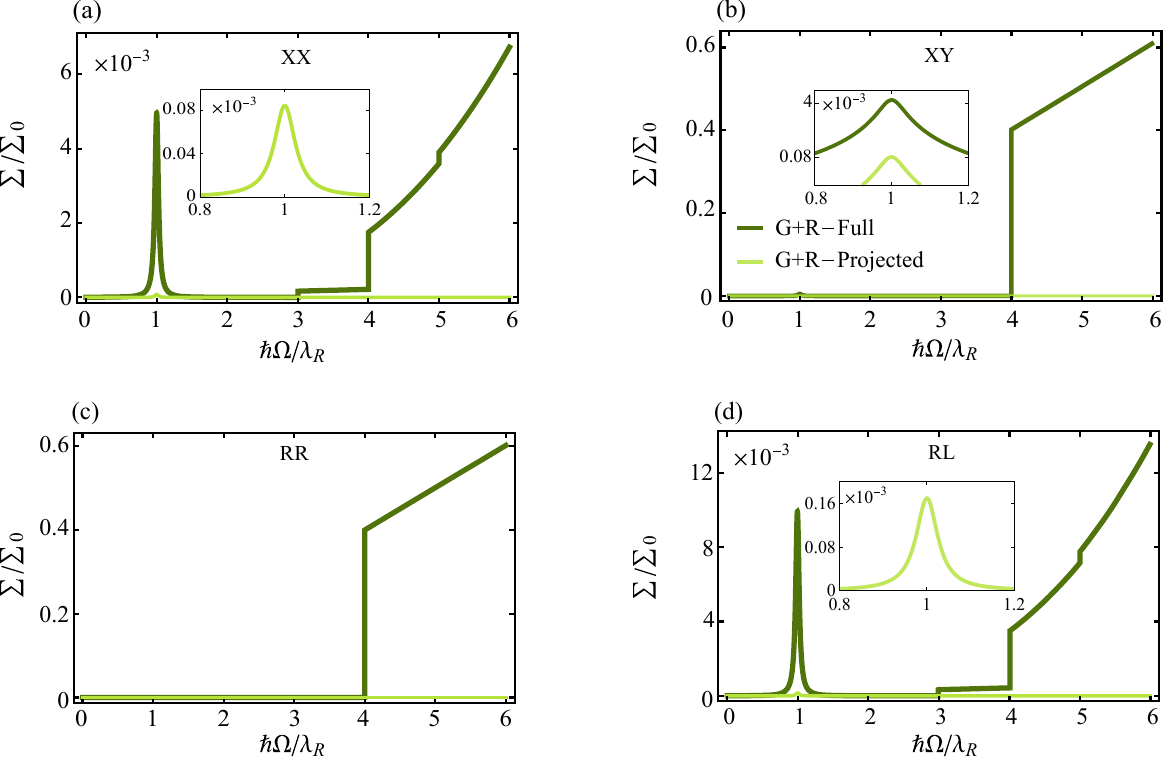}
    \caption{Comparison of differential cross-sections for graphene with Rashba SOC calculated in the $4\times4$ Hilbert space (dark) and in the projected space (light) with the $\mu$ in the conduction bands. The inset of panel (b) uses a logarithmic scale on the y‑axis. The model in the projected space can only capture the sharp spin-flip excitations of the low-energy Hilbert space, but it does so with a very weak signal strength. This indicates that models used to predict Raman signals will get the right resonant frequencies, but a significantly weaker signal strength as it misses significant spectral weight from the intermediate transitions to the off-shell (non-resonant) states from other bands. Here, $\mu=2\lambda_{\rm R}$, $\hbar\Omega_{\rm I}=20\lambda_{\rm R}$, and $\lambda_{\rm R}\tau/\hbar\sim 33$.}
    \label{fig:diff_cross_ProjG_Comp}
\end{figure}

Thus, we note that while working in the projected systems allows us to get the correct low-energy excitations and its symmetries, it misses significant scattering amplitude even for excitations that only involve the low-energy states. In Fig. \ref{fig:diff_cross_ProjG_Comp} we compare the differential cross-sections computed from our prescription in all scattering geometries for the $4\times4$ and projected Hilbert spaces of mono-layer graphene. As expected, we see a noticeably small cross-section calculated in the projected subspace, indicating that one can get significant enhancement of the signal from the off-shell states of the bands projected out.

\section{Comparison with a 2D electron gas}\label{Sec:Comparison_2DEG}
Another fitting comparison to make is between the responses of the projected graphene system with Rashba SOC with a 2DEG system with Rashba SOC. This is because at the level of the $2\times2$ low-energy bands, both display two conduction bands with chiral and split Fermi surfaces. The Hamiltonian and the associated perturbations for the 2DEG are
\bse
\bea
\hat{\mathcal{H}}^{\rm 2DEG,R}_0&=&\frac{\mathbf{p}^2}{2m}\hat s^0+v_{\rm R}(\mathbf{p}\times\hat {\mathbf s})\cdot\hat z,\label{eq:model2DEGRa}\\
\hat{\mathcal{H}}^{\rm 2DEG,R}_A&=&\frac{e^2}{2m}\hat{\bA}^2(\br,t)\hat{s}^0+\frac{\bp}{m}\cdot e\hat{\bA}(\br,t)\hat{s}^0+v_R~  e\hat{\bA}(\br,t)\times\hat{\mathbf s},\label{eq:model2DEGRb}
\eea
\ese
where $m$ is the band mass of the 2DEG and $v_{\rm R}$ is the Rashba coupling with units of velocity. Due to the presence of $\bp^2$, we also have to keep the direct-scattering contribution. Apart from the multivalley nature, the chirality of the eigenstates and the Fermi surface splitting of the 2DEG systems is similar to that of graphene. Despite this, the nature of coupling to photons via the $\hat\bA$ field is different. This will lead to important differences in cross-section. To see this, we can again employ Eq. \eqref{eq:NewMFi} to compute the Raman tensors in the space $\{|cl\rangle,|cr\rangle\}$:
\begin{align}\label{eq:2D M}
\hat m^{\rm D}_{xx}(\theta_p)&=\frac{m_e}m\begin{pmatrix}
    1&0\\
    0&1
\end{pmatrix},\nn\\
\hat m^{\rm L}_{xx}(\theta_p)&=0,\nn\\
\hat m^{\rm SL}_{xx}(\theta_p)&=-\frac{2m_ev_{\rm R}^2v^{\rm 2D}_{\rm F}p}{(\hbar\Omega_{\rm I})^2}\begin{pmatrix}
    -\frac{2v_{\rm R}}{v^{\rm 2D}_{\rm F}}\sin^2\theta_p&-i\left(\frac{p-mv_{\rm R}}{mv^{\rm 2D}_{\rm F}}\right)\sin2\theta_p\\
    -i\left(\frac{p+mv_{\rm R}}{mv^{\rm 2D}_{\rm F}}\right)\sin2\theta_p&\frac{2v_{\rm R}}{v^{\rm 2D}_{\rm F}}\sin^2\theta_p
\end{pmatrix},\nn\\~\nn\\
\hat m^{\rm D}_{yy}(\theta_p)&=\hat m^{\rm D}_{xx}(\theta_p),\nn\\
\hat m^{\rm L}_{yy}(\theta_p)&=0,\nn\\
\hat m^{\rm SL}_{yy}(\theta_p)&=\hat m^{\rm SL}_{xx}(\theta_p+\pi/2),\nn\\~\nn\\
\hat m^{\rm D}_{xy}(\theta_p)&=0,\nn\\
\hat m^{\rm L}_{xy}(\theta_p)&=-\frac{m_ev_{\rm R}^2}{\hbar\Omega_{\rm I}}\begin{pmatrix}
    0&i\\
    i&0
\end{pmatrix},\nn\\
\hat m^{\rm SL}_{xy}(\theta_p)&=-\frac{2m_ev_{\rm R}^2v^{\rm 2D}_{\rm F}p}{(\hbar\Omega_{\rm I})^2}\begin{pmatrix}
    \frac{v_{\rm R}}{v^{\rm 2D}_{\rm F}}\sin2\theta_p&i\left(\frac{p-mv_{\rm R}}{mv^{\rm 2D}_{\rm F}}\right)\cos2\theta_p\\
    i\left(\frac{p+mv_{\rm R}}{mv^{\rm 2D}_{\rm F}}\right)\cos2\theta_p&-\frac{v_{\rm R}}{v^{\rm 2D}_{\rm F}}\sin2\theta_p
\end{pmatrix},\nn\\~\nn\\
\hat m^{\rm D}_{yx}(\theta_p)&=0,\nn\\
\hat m^{\rm L}_{yx}(\theta_p)&=-\hat m^{\rm L}_{xy}(\theta_p),\nn\\
\hat m^{\rm SL}_{yx}(\theta_p)&=\hat m^{\rm SL}_{xy}(\theta_p).
\end{align}
Comparing this result with the multi-valley case (projected case), we observe that the relations between the tensor components are the same. However, there are two differences in the structure of the tensors. One is the presence of the direct term $\hat m^{\rm D}$. Since this is only non-trivial for intra-band excitations, they will not affect results at finite Raman shifts. The other difference is that the leading term in the XY geometry in the case of 2DEG correspond to \textit{chirality-flipping} excitations between states $|cl\rangle,|cr\rangle$, when it \textit{preserved chirality} in the $4\times4$ Hilbert space of graphene. In fact, in the latter case, the excitations were between the valence and conduction bands. In the projected space, we don't have any leading excitations, which is consistent with the fact that we do not have the valence bands in the model. This difference in XY geometry also manifests in the RR geometry which is now sensitive to chiral excitations, when such excitations were forbidden in the case of graphene. Nevertheless, the subleading contributions have the same anisotropy as the (projected) multi-valley case with their values even matching when $v_{\rm R}\rightarrow\lambda_{\rm R}/2p$ and $p\gg mv_{\rm R}$. This is reflective of the indistinguishable nature of the spin-split Fermi surfaces in the two cases. 

Calculating the differential cross-section we get
\bea\label{eq:2dR_diff2}
\frac{d^2\sigma}{d\mathcal Od\Omega}&=&\frac{r_0^2}{\Omega_{\rm I}}\frac{S(m_ev^{\rm 2D}_{\rm F})^2}{\pi\hbar^2}\left[\frac{\Omega}{2\Omega_{\rm I}}\frac{\lambda_{\rm R}^2}{\mu^2}\Xi_{\rm L}+\frac{\Omega^3}{16\Omega^3_{\rm I}}\frac{(\lambda_{\rm R}^2+2\mu\hbar\Omega)^2}{\mu^2\lambda_{\rm R}^2}\Xi_{\rm SL}\right]\Theta(E^2_-/4\mu<\hbar\Omega<E^2_+/4\mu)\nn\\
\text{where }E^2_{\pm}&=& \lambda_{\rm R}\sqrt{\lambda_{\rm R}^2+16\mu^2}\pm\lambda_{\rm R}^2,~~\lambda_{\rm R}=2v_{\rm R} p_{\rm F}.
\eea
Here, $v_F^{\rm 2D}$ is the Fermi velocity of the 2DEG at Fermi energy $\mu$ and $p_F$ is the associated Fermi momentum. When comparing the result with graphene, we note that the overall prefactor $\propto$ the square of the Dirac velocity in graphene and to the square of the Fermi velocity in lightly doped/gated semiconductors. The latter can be $\sim100$ times smaller than the Dirac velocity. This can reduce the scattering cross-section for a 2DEG by a factor $\sim10^4$. Further, the prefectors to the polarization factors $\Xi_{\rm L},\Xi_{\rm SL}$ also contain additional factors $\propto(\lambda_{\rm R}/\mu)^2$, when compared to similar terms in graphene [see e.g. Eq. \eqref{GVZ_diff}], which can further reduce the signal strength for small SOC. Thus, graphene and graphene-like systems offer a better opportunity to observe the SOC effects in a Raman probe. The improvement is due to the off-shell contributions from the higher energy states in the full Hilbert space and the large Dirac velocity.

\begin{figure}[htp]
    \hspace*{-2cm}
\centering
    \includegraphics[width=0.9\linewidth]{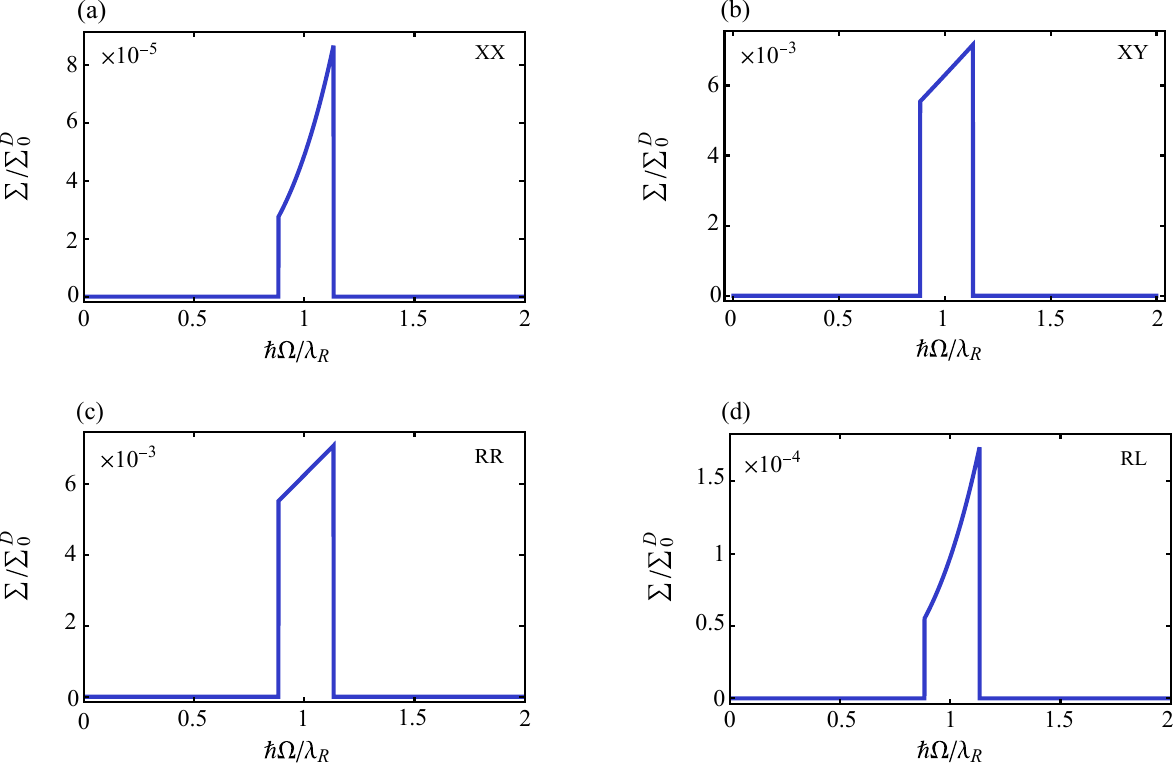}
    \caption{The normalized differential cross-section of a 2DEG system for various polarizations. Here $\Sigma_0^{\rm D}$ is given by the same expression as in Fig. \ref{fig:diff_cross_G_Rashba} but with $v_F$ being that of the 2DEG. The spread in the peak is due to broadened nature of the chirality-flip excitations which in turn is due to the momentum dependent splitting of the electronic structure in the 2DEG. While all channels pick up the chiral excitations, the signal is the strongest in XY and RR. Here $\mu=2\lambda_{\rm R}$.}
    \label{fig:2D-Rashba}
\end{figure}

In Fig. \ref{fig:2D-Rashba} we plot the differential cross-section for a 2DEG for various scattering geometries. As in the case of graphene, the signal in the XY and RR geometries are the strongest, but unlike graphene, the entire signal arises from chirality flipping excitations. We caution the reader that the normalization in Fig. \ref{fig:2D-Rashba} involves the Fermi velocity of the 2DEG and hence is smaller than the normalization constant of graphene. The characteristic difference in the response form the (projected) graphene and 2DEG systems despite having the same type of SOC and the same number of states participating in the response, highlights the importance of the Hilbert space where SU(2) is broken.

\section{Conclusions and experimental outlook}\label{Sec:Conclusion}
This work pursued two objectives. First, to understand how SOC arising from inversion breaking affects light matter coupling and second, to study variances in the nature of this coupling across different systems. Regarding the first point, we have seen that the components of SOC that couple the electron spin to the momentum quantum number (such as Rashba) also couple spin-flip excitations to photon scattering, albeit, with some selectivity with respect to the polarization geometry. This coupling is different from the one that emerges in resonant Raman scattering which exists even without SU(2) symmetry breaking in the Hilbert space and hence (potentially) observable without tuning the laser to an internal resonance of the system.

The details of the selectivity depends on the spin structure of the Hilbert space. This was the aim of the second objective. We studied 3 models to elucidate the differences. The first two were for a SOC graphene system. One with a $4\times4$ Hilbert space formed out of spin and sublattice degrees of freedom, and the other a $2\times2$ subspace formed by projecting on to the states involving the Fermi surface. In both these models, we capture the low-energy chiral-spin resonance between the spin-split states at the Fermi surface in all polarization geometries except RR. However, the signal is significantly amplified in the $4\times4$ case due to the presence of contributions from the intermediate bands. The strongest signal is in XY and RR geometries where the leading contributions are, in fact, from non-spin-flip excitations. The other model we considered was for a 2DEG, where the spin-flip resonances were observed in all polarizations geometries. The strongest signals remain in XY and RR geometries, but unlike graphene, they are entirely due to spin-flip excitations. This difference highlights the relevance of the Hilbert space itself to the modelling of light-matter coupling.

Finally, it is befitting to comment on the prospects of experimental consequences of these effects. First, a distinction between VZ-dominated and Rashba-dominated case can be drawn by fitting the line-shape at higher energies (see Fig. \ref{fig:diff_cross_G_RashbaVZ}). Second, if low-energy peak is detectable in XX,XY and RL, then the presence or absence of a peak in RR can serve as a distinguishing feature of contributions from a Dirac-like or a parabolic 2DEG-like low-energy Hilbert space. 

The above information is deduced from the line-shape of the spectrum. One can now estimate if the above line shapes themselves are detectable in an experiment. To this effect, consider the frequency integrated cross-section over a range of frequencies $\Omega_{\rm min}$-$\Omega_{\rm max}$
\beq\label{eq:FreqInt}
\frac{d\sigma}{d\mathcal O}=\int_{\Omega_{\rm min}}^{\Omega_{\rm max}}d\Omega \frac{d^2\sigma}{d\mathcal Od\Omega}.
\eeq
To check if the predicted features are detectable, we can make an estimate for the order of magnitude of the differential cross-section for various terms for the $4\times4$ model for graphene by using the $\Xi_{\rm L},\Xi_{\rm SL}$ components from Eq. \eqref{GVZ_diff}, and the $s_3$ component from Eq. \eqref{eq:GR_diff} (for the resonance):
\begin{align}
    \frac{d\sigma}{d\mathcal O}=r_0^2S(m_ev_F/\hbar)^2/\pi\times\begin{cases}
\left(\frac{\Omega_{\rm max}}{2\Omega_{\rm I}}\right)^2,~\text{Leading contribution (no SOC)},\\
\frac14\left(\frac{\Omega_{\rm max}}{2\Omega_{\rm I}}\right)^4,~\text{Subleading contribution (SOC correction)}\\
\frac{\lambda_{\rm R}^3\mu}{(\hbar\Omega_{\rm I})^4},~\text{Chiral-spin resonance contribution}.
    \end{cases}
\end{align}
In graphene $v_F\sim 10^6$ms$^{-1}$. Typical Raman lasers have $\hbar\Omega_{\rm I}\sim2$eV, gating up to $\mu\sim100$meV is reasonable in these systems~\cite{Yoon2013,Gallais2016}, and typical SOC in proximity induced graphene on TMDs~\cite{morpurgo:prx} is $\lambda_{\rm R}\sim 10$meV. Using these numbers, and setting $\Omega_{\rm min}=0$, $\hbar\Omega_{\rm max}=4\mu=400$meV, along with the choice of a circular sample of radius $\tilde r_s$ nm, we get  
\begin{align}\label{eq:nums}
    \frac{d\sigma}{d\mathcal O}\sim \tilde r_s^2\times10^{-28}{\rm m}^2{\rm sr}^{-1}\times\begin{cases}
6\times 10^{-2},~\text{Leading contribution (no SOC)},\\
1.5\times 10^{-4},~\text{Subleading contribution (SOC correction)}\\
4\times 10^{-8},~\text{Chiral-spin resonance contribution}.
    \end{cases}
\end{align}
Signals up to $10^{-34}{\rm m}^2{\rm sr}^{-1}$ are detectable in Raman scattering. Equation \eqref{eq:nums} suggests that the SOC-blind $2\mu$ feature easily falls in the detectable range. Indeed this has been detected in Refs. ~\cite{Gallais2016, Streltsov2015}. The effects of SOC at higher energies also falls in the detectable range (for the parameters specified above). The strength of the chiral-spin resonance, however, is below the threshold. The above signal strength is calculated for 1 nm radius sample and can certainly be boosted for larger size (the dimensionless factor of $\tilde r_s^2$). As shown in Sec. \ref{Sec:Projection}, the spectral weight may also get a boost from other higher energy intermediate states in graphene or graphene-like systems, but this cannot be reliably estimated within the above model. Effects of electronic correlations, which induce vertex corrections to the final state, can further alter the overall cross-section, but that estimate is beyond the scope of this work. 

\begin{figure}[htp]
\centering
    \includegraphics[width=\linewidth]{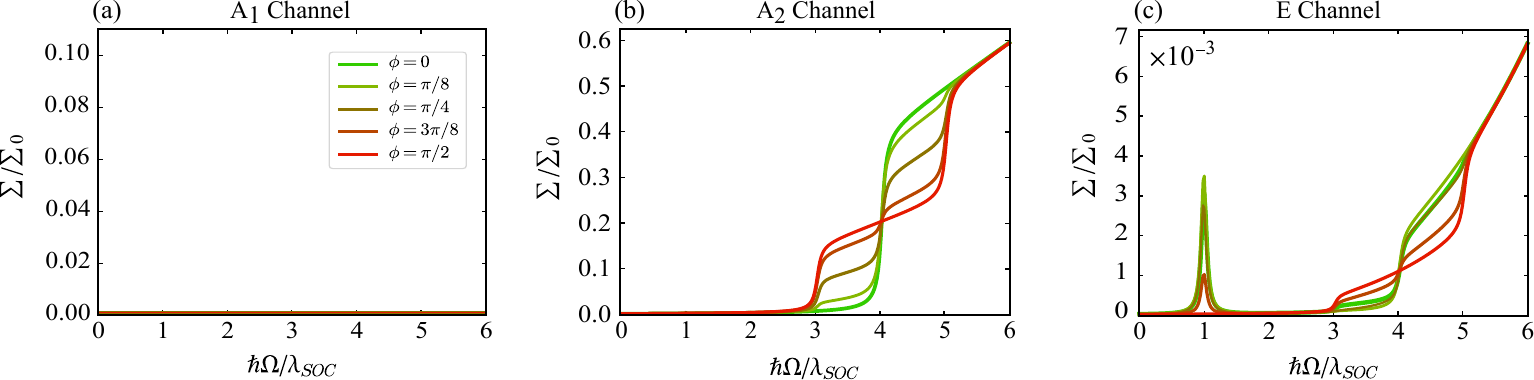}
    \caption{Polarized Raman signals from Fig. \ref{fig:diff_cross_G_RashbaVZ} decomposed into the IRREPs of C$_{3v}$. The fully symmetric channel $A_1$ picks up nothing, the anti-symmetric channel $A_2$ picks up the $2\mu$ jump and split features. This contribution arises from the leading contribution $\Xi_{\rm L}$. The vector channel $E$ picks up the subleading contributions from $\Xi_{\rm SL}$ which contains the resonance and some additional features around $2\mu$.}
    \label{fig:IRREPs}
\end{figure}

We take this opportunity to place the above proposed Raman features in the context of other Raman active excitations in graphene. In clean monolayer, there is a strong G-phonon (of $E$ symmetry) signature at $\sim 200$ meV with a linewidth$\sim 10$ meV. Beyond this are the 2D phonons (of $A_1$-symmetry) at $\sim 330$meV. The spin resonances are of the order of $10$s of meV and hence well separated from these features. The leading contribution which captures the split in the step at $2\mu$, on the other hand, can interfere with these Raman active modes. One easy way to avoid this is to tune $2\mu$ (e.g. by gating) to be away from these frequencies. Even if they are close, one can use the polarization-resolved signals to remove the phonon contributions. The 2D phonon line has $A_1$ symmetry, the $G$ line has $E$ symmetry\footnote{The $G$-line is known to interact with graphene's electron-hole continuum, but this interaction is weakened due to screening from the high doping~\cite{Yoon2013}. The weaker this interaction is, the weaker is the possibility of cross-leakage into other channels.} and the leading term proposed above has $A_2$ symmetry~\cite{Kashuba2009,Gallais2016}. We show the symmetry decomposed contributions in Fig. \ref{fig:IRREPs} where the two contributions, namely the leading and subleading ones, couple to the $A_2$ and the $E$ channels, with the former being significantly stronger. See Appendix \ref{App:graphene_symmetry} for the decomposition details.

Finally, it should also be noted that the cross-sections for 2DEG and even the projected graphene models can be smaller by $\sim10^{-4}$ for various reasons described in the text. This means that one must take care in modeling the response when using only the low-energy subspace as the estimates for the signals can be artificially weak. Thus, despite a weak signal strength prediction from low-energy model based theory, an experimental verification is still encouraged.

The scheme outlined in this work serves as a stepping stone in the direction of modeling light matter coupling in a wide variety of systems where novel low-energy \textit{electronic} properties emerge from symmetry breaking, engineering (such us Van der Waals structures and their twisted counterparts), or even phase transitions. We also anticipate a pursuit of studies of electronic phenomena in systems where previous studies primarily focused only on phonon excitations (e.g. in $\rm MoS_2$ ~\cite{Lee2010}).

\section{Acknowledgments} 
The authors thank G. Blumberg for several illuminating discussions and offering consistency checks for our results. This work was funded by the Natural Sciences and Engineering Research Council of Canada (NSERC) Grant No. RGPIN-2019-05486. 

\appendix
\section{General derivation of the differential cross-section}\label{Sec:App1}
\textit{A note on convention}: Here we will follow a convention where $\bf{x}$ will denote a 2D vector that lies in the plane of the 2D sample, while $\vec x$ would denote a 3D vector. Further, The modulus of $\bf x$ or $\vec x$ would be denoted by $x$.

We begin with a Hamiltonian that is analytic in $\bp$ in the low-energy sector. We couple photons through the minimal coupling prescription where we promote the matrix elements $ \mathcal{H}_{ab}(\bp)\rightarrow \mathcal{H}_{ab}[\bp+e\hat{\bA}(\vec r,t)]$, where the vector potential is given by\footnote{This is the standard definition in SI units. We have chosen to express it in a form that makes it explicit that the dimension of $e\bf A$ is that of momentum.}
\beq\label{eq:VecPotDef}
\hat\bA(\vec r,t)\equiv\sum_{\vec q}
\frac{\hbar}e\sqrt{\frac{2\pi\alpha}{Vq}}
\left[e^{i\{\vec q\cdot\vec r-\Omega(q) t\}}~\bl~\hat b_{\vec q,\bl}+h.c.\right].
\eeq
Here $\alpha$ is the fine structure constant, $V$ is the volume of the sample over which the interaction with the electromagnetic field takes place, $\hat b_{\vec q,\bl}$ is the annihilation operator of a photon with wavenumber $\vec q$ and polarization $\bl$ (which can be complex), and $\Omega(q)$ is the angular frequency of the above photon given by \beq\label{eq:phton_disp}\Omega(q)=cq=c\sqrt{{\bf q}^2 + q_z^2},\eeq where $c$ is the speed of light. A popular geometry for Raman scattering setup is a near-normal incidence of the electromagnetic wave on the sample. This means that the oscillatory electric and magnetic field components lie in the plane of the 2D sample. For this reason, we will have $\partial_\br$ and $\hat \bA$ (and hence $\bl$) to remain in the 2D plane. The position dependence of the $\hat \bA$ field itself could have a 3D $\vec r$ dependence. This is also why the mode-expansion of $\bA$ is made in terms of the 3D $\vec q$ vectors. The matrix elements $\mathcal H_{ab}$ correspond to the Hamiltonian $\hat H = \mathcal H_{ab}\hat c^\dagger_a\hat c_b$, where $\hat c_b$ is the annihilation operator for state $b$.

The light-matter interaction terms are $\propto\bA$ and the ones relevant for Raman scattering are presented in \cref{eq:Hint_Def}. Raman scattering counts the number of photons scattered per unit time into a solid angle $d\mathcal O$ (measured from the sample), and in energy between the scattered photon frequency $\Omega_{\rm S}$ and $\Omega_{\rm S}+d\Omega$. An intrinsic property of this scattering is the differential cross-section (which is the above count divided by the incoming flux of photons), which can be found by calculating the transition between the states of the photon+sample system from an initial state 
$$|{\rm I}\rangle\equiv |1_{\vec q_{\rm I}},0_{\vec q_{\rm S}}\rangle|i\rangle$$ 
to a final state
$$|{\rm F}\rangle\equiv |0_{\vec q_{\rm I}},1_{\vec q_{\rm S}}\rangle|f\rangle,$$
where $\vec q_{\rm I},\vec q_{\rm S}$ are the wavenumbers of the incident and scattered photons, $|x\rangle$ for $x\in\{i,f\}$ are the initial and final electronic eigenstates of the sample (in the absence of the photon field). This transition matrix element can be computed using perturbation theory with the perturbation provided by $\hat{\mathcal H}_A$. Since $\hat A_{\alpha}$ can only annihilate or create a single photon, moving the occupancy of photons from one to the other requires a second order process in $\hat A_{\alpha}$. This means that we have to collect all terms in the perturbation series that yields $\hat A_{\alpha}\hat A_{\beta}$ terms. This happens already at first order perturbation theory for the diamagnetic term, which we call the direct scattering. The only other source of $\hat A_{\alpha}\hat A_{\beta}$ is the second order contribution from the absorption vertex. 

Accounting for all these terms we can arrive at the dimensionless two-photon scattering matrix element $M_{fi}$ (Kramer-Heisenberg Formula ~\cite{Kramers1925,sakurai1967})
\bea\label{eq:Mfi}
M_{fi}=\underbrace{-m_e\sum_\nu\left[\frac{\langle f|\partial_{p_{\alpha}}\hat{H}_0 \ell_{\alpha}^{\rm S*}e^{-i\vec q_{\rm S}\cdot\vec r}|\nu\rangle\langle \nu|\partial_{p_{\alpha}}\hat{H}_0 \ell_{\alpha}^{\rm I}e^{i\vec q_{\rm I}\cdot\vec r}|i\rangle}{E_\nu-\hbar\Omega_{\rm I}-E_i}+\frac{\langle f|\partial_{p_{\alpha}}\hat{H}_0 \ell_{\alpha}^{\rm I}e^{i\vec q_{\rm I}\cdot\vec r}|\nu\rangle\langle \nu|\partial_{p_{\alpha}}\hat{H}_0 \ell_{\alpha}^{\rm S*}e^{-i\vec q_{\rm S}\cdot\vec r}|i\rangle}{E_\nu+\hbar\Omega_{\rm S}-E_i}\right]}_{\rm indirect}&&\nn\\
+~\underbrace{m_e\ell_{\alpha}^{\rm S*}\ell_{\beta}^{\rm I}\langle f|e^{i(\vec q_{\rm I}-\vec q_{\rm S})\cdot\vec r}\partial_{p_{\alpha}}\partial_{p_{\beta}}\hat{H}_0|i\rangle}_{\rm direct}&&,
\eea
where, $\nu$ is a label for intermediate electronic states of the system with energy $E_\nu$, $\hat H_0\equiv \hat c^{\dagger}_{a}\mathcal{H}_{0,ab}\hat c_b$ (repeated indices are summed over), $\partial_{p_{\alpha}}\hat H_0\equiv \hat c^{\dagger}_{a}\partial_{p_{\alpha}}\mathcal{H}_{0,ab}\hat c_b$, and $\partial_{p_{\alpha}}\partial_{p_{\beta}}\hat H_0\equiv \hat c^{\dagger}_{a}\partial_{p_{\alpha}}\partial_{p_{\beta}}\mathcal{H}_{0,ab}\hat c_b$, where $\hat c^\dagger_a$ is the creation operator for some fermionic basis states $\{a\}$. Due to the presence of the fermionic creation/annihilation operators, $\langle f|..|i\rangle$ leads to the occupancy factor of $n_{\rm F}(E_i)[1-n_{\rm F}(E_f)]$ requiring the initial state to be occupied and the final state to be empty. The first term is the indirect two-photon scattering, while the second one is the direct two-photon scattering. For any generic low-energy Hamiltonian, Eq. \eqref{eq:Mfi} for $M_{fi}$ can be used by replacing $\partial_\alpha\hat H$ and $\partial_\alpha\partial_\beta\hat H$ with the coefficients of $\hat A_{\alpha}$ and $\hat A_{\alpha}\hat A_{\beta}$ in $\mathcal H_A$. Note that the $\hat b$ operators got used up in contracting the photon states and hence don't appear explicitly in the formula for the scattering matrix element. The sum over the intermediate states $|\nu\rangle$ run over all states of the system: all bands and all $\bk$ vectors of the Brillouin Zone (BZ). We can factor the electronic wavefunction into a Hilbert-space component and a plane-wave component such that $|x\rangle=|u_x\rangle\sqrt{\frac{\kappa}{A}}e^{i\bk_x\cdot\br}e^{-\kappa|z|}$, where we have used the fact that our states lie in a 2D plane of area $A$, and $\kappa$ captures the length scale associated with the spread of the wavefunction in the $z$-direction. The matrix elements that appear in $M_{fi}$  will then involve real space integrations of the type: 
\bea\label{eq:realspaceint}
\langle f|e^{i\vec q\cdot\vec r}Z|x\rangle&=&\frac{\kappa}A\int_{\br}e^{-i(\bk_f-\bk_x-\bq)\cdot\br}\int_{-\infty}^\infty dz~e^{-2\kappa|z|+iq_zz}\langle\phi_f|Z|\phi_x\rangle\nn\\
&=&\delta_{\bk_f-\bk_x,\bq}~\frac{4\kappa^2}{4\kappa^2+q_z^2}\langle\phi_f|Z|\phi_x\rangle.
\eea
We next observe that $\bk$'s are fermionic momenta in the BZ that are $\sim {\rm nm}^{-1}$ and $\bq,q_z$ are the wavevector components of the photon that are $\sim 0.002{\rm nm}^{-1}$ which are $\ll \bk$'s. Thus, all the matrix elements are restricted to $\delta\bk\approx0$. Physically, this means that only `vertical' transitions happen in the band structure. This restricts the sum over intermediate states $|\nu\rangle$ to only include vertically stacked states (at the the same $\bk$). We also expect $\kappa\sim {\rm \AA}^{-1}$ which allows us to ignore $q_z$ in relation to $\kappa$. Going forward, we shall denote $\langle f|e^{i\vec q\cdot\vec r}Z|x\rangle\rightarrow\langle f|Z|x\rangle\delta_{\bk_f,\bk_x}$, with the understanding that the matrix elements are now only in the matrix structure of the Hilbert space, with the position part already integrated out.

We comment here on the size of the space over which $\nu$ should be summed. First note that a completeness relation can be defined over any subspace $\{|\nu\rangle\}$ of states of the Hamiltonian (that are not degenerate with states outside of this space), provided all calculations only involve this subspace. This is where it becomes important to choose the basis states in which the minimal coupling is performed. Within this basis, one can rest assured that the vector potential would only couple states within this subspace. This effectively is a restriction in the magnitude of $\vec q$ of the photon states that couple to our system. To couple to very high-energy states, we would have to extend the range of the photon states and there it will become essential to expand the Hilbert space again.

The above considerations allow us to arrive at Eq. \eqref{eq:Mfi2}. In the non-resonant case, if we use a laser such $\Omega_{\rm I}\gg{\rm Max}[E_\alpha]$ (but lower than intermediate states from other subspaces), then indirect contributions could be Taylor-expanded in $(E_\nu-E_i)/\hbar\Omega_{\rm I/S}$. Consider the first term of the indirect process. The leading and subleading contributions can be written as 
\bea\label{eq:math1}
\sum_\nu\frac{\langle f|\partial_{p_{\alpha}}\hat{H}_0 \ell_{\alpha}^{\rm S*}|\nu\rangle\langle \nu|\partial_{p_{\alpha}}\hat{H}_0 \ell_{\alpha}^{\rm I}|i\rangle}{E_\nu-\hbar\Omega_{\rm I}-E_i}&\approx& \sum_\nu\frac{\langle f|\partial_{p_{\alpha}}\hat{H}_0 \ell_{\alpha}^{\rm S*}|\nu\rangle\langle \nu|\partial_{p_{\alpha}}\hat{H}_0 \ell_{\alpha}^{\rm I}|i\rangle}{-\hbar\Omega_{\rm I}}\left[1+\frac{E_\nu-E_i}{\hbar\Omega_{\rm I}}\right]\nn\\
&=&-\frac{\langle f|\partial_{p_{\alpha}}\hat{H}_0 \ell_{\alpha}^{\rm S*}\partial_{p_{\beta}}\hat{H}_0 \ell_{\beta}^{\rm I}|i\rangle}{\hbar\Omega_{\rm I}}+\frac{\langle f|\partial_{p_{\alpha}}\hat{H}_0 \ell_{\alpha}^{\rm S*}[\partial_{p_{\beta}}\hat{H}_0,\hat{H}_0] \ell_{\beta}^{\rm I}|i\rangle}{\hbar^2\Omega^2_{\rm I}}\nn\\
&=&-\left[\frac{\langle f|\partial_{p_{\alpha}}\hat{H}_0 \partial_{p_{\beta}}\hat{H}_0 |i\rangle}{\hbar\Omega_{\rm I}}-\frac{\langle f|\partial_{p_{\alpha}}\hat{H}_0 [\partial_{p_{\beta}}\hat{H}_0,\hat{H}_0] |i\rangle}{\hbar^2\Omega^2_{\rm I}}\right]\ell_{\alpha}^{\rm S*}\ell_{\beta}^{\rm I}
\eea
Similarly, the second term of the indirect scattering yields
\bea\label{eq:math2}
\sum_\nu\frac{\langle f|\partial_{p_{\alpha}}\hat{H}_0 \ell_{\alpha}^{\rm S*}|\nu\rangle\langle \nu|\partial_{p_{\alpha}}\hat{H}_0 \ell_{\alpha}^{\rm I}|i\rangle}{E_\nu+\hbar\Omega_{\rm S}-E_i}&\approx& -\left[\frac{\langle f|\partial_{p_{\beta}}\hat{H}_0 \partial_{p_{\alpha}}\hat{H}_0 |i\rangle}{-\hbar\Omega_{\rm S}}-\frac{\langle f|\partial_{p_{\beta}}\hat{H}_0 [\partial_{p_{\alpha}}\hat{H}_0,\hat{H}_0] |i\rangle}{\hbar^2\Omega^2_{\rm S}}\right]\ell_{\alpha}^{\rm S*}\ell_{\beta}^{\rm I}.
\eea
Combining these results, we find that
\bea\label{eq:NewMFiApp}
M_{fi}&=&\langle f|\hat M_{\alpha\beta}|i\rangle\ell_{\alpha}^{\rm S*}\ell_{\beta}^{\rm I}~\delta_{\bk_f,\bk_i},\nn\\
\text{such that }\frac{\hat M_{\alpha\beta}}{m_e}&=&\underbrace{\frac{\partial_{p_{\alpha}}\hat H_0\partial_{p_{\beta}}\hat H_0}{\hbar\Omega_{\rm I}}-\frac{\partial_{p_{\beta}}\hat H_0\partial_{p_{\alpha}}\hat H_0}{\hbar\Omega_{\rm S}}}_{\rm leading~indirect}\underbrace{-\frac{\partial_{p_{\alpha}}\hat H_0[\partial_{p_{\beta}}\hat H_0,\hat H_0]}{\hbar^2\Omega^2_{\rm I}}-\frac{\partial_{p_{\beta}}\hat H_0[\partial_{p_{\alpha}}\hat H_0,\hat H_0]}{\hbar^2\Omega^2_{\rm S}}}_{\rm subleading~ indirect}+\underbrace{\partial_{p_{\alpha}}\partial_{p_{\alpha}}\hat H_0}_{\rm direct}.
\eea
When working with a low-energy subspace obtained from Hilbert space projections, we simply replace $\partial_{p_\alpha}\hat{\mathcal H}$ and $\partial_{p_\alpha}\partial_{p_\beta}\hat{\mathcal H}$ with $\hat v_\alpha$ and $\hat I_{\alpha\beta}$ as obtained from the projection procedure. 

\section{Symmetry decomposition of graphene}\label{App:graphene_symmetry}
Here we re-express the XX,XY,RL, and RR signals in terms of the irreducible representations (IRREPs) of graphene's Hamiltonian at the K/K' points. Given the Hamiltonian in Eq. \eqref{eq:modelG4Rb} the applicable symmetry group is C$_{3v}$. In this group we have the following relations:
\begin{table}[htp]
    \centering
    \begin{ruledtabular}
    \begin{tabular}{cc}
    Raman polarization geometry & IRREPs of C$_{3v}$\\
    \hline\hline
    XX (=YY) & $A_1+E$\\
    XY & $A_2+E$\\
    RL & $2E$\\
    RR & $A_1+A_2$
    \end{tabular}
    \end{ruledtabular}
    \caption{Relations between the signals collected in various Raman polarizations and their IRREP contributions}
    \label{tab:sym_decomp}
\end{table}
This can be used to infer the signals in individual IRREPs: $A_1=XX-RL/2$, $E=RL/2$, and $A_2=XY-RL/2$. For the Rashba case, working with the polarization factors $\Xi_{\rm L}$ and $\Xi_{\rm SL}$, we infer from Eq. \eqref{eq:DiffXSec2} that the signals in $A_1=0$, $E\propto \Xi_{\rm SL}$, and $A_2\propto \Xi_{\rm L}$. That is, the $A_2$ channel primarily captures the $2\mu$ features, whereas the $E$ channel captures the spin resonance. In Fig. \ref{fig:IRREPs} we show the symmetry decomposed result of Fig. \ref{fig:diff_cross_G_RashbaVZ} which validates the above statement. As stated in Sec. \ref{Sec:Conclusion}, in the Raman spectroscopy of graphene, the G-phonon interference is expected in the E-channel at $\sim 200$meV. Thus, the spin resonance feature, though weak, is decoupled from it. The stronger $A_2$ feature, which hosts valley-Zeeman dominated splittings is also decoupled from the E-channel G-phonon.

\bibliography{References_main.bib}
\end{document}